\RequirePackage{fix-cm}
\documentclass[twocolumn,natbib]{svjour3}          
\smartqed  
\usepackage{graphicx}
\usepackage{amsmath}
\usepackage{amssymb}
\usepackage{subfig}
\usepackage{tabularx}

\newcommand{\degree}{\ensuremath{^{\circ}}} 
\newcommand{\rmsub}[1]{\ensuremath{_{\mathrm{#1}}}} 

\newcommand{\secref}[1]{Sec.~\ref{sec:#1}}

\newcommand{\figref}[1]{Fig.~\ref{fig:#1}}
\newcommand{\figsref}[1]{Figs.~\ref{fig:#1}}

\newcommand{\tabref}[1]{Table~\ref{tab:#1}}

\newcommand{\eqnrefjustpar}[1]{(\ref{eq:#1})}
\newcommand{\eqnref}[1]{Eq.~\eqnrefjustpar{#1}}
\newcommand{\eqnsref}[1]{Eqs.~\eqnrefjustpar{#1}}

\newcommand{\knI}{\ensuremath{k\rmsub{nI}}}
\newcommand{\knP}{\ensuremath{k\rmsub{nP}}}
\newcommand{\kf}{\ensuremath{k\rmsub{f}}}
\newcommand{\thetan}{\ensuremath{\theta\rmsub{n}}}
\newcommand{\thetaf}{\ensuremath{\theta\rmsub{f}}}
\newcommand{\thetandot}{\ensuremath{\dot{\theta}\rmsub{n}}}
\newcommand{\thetafdot}{\ensuremath{\dot{\theta}\rmsub{f}}}
\newcommand{\Pp}{\ensuremath{P\rmsub{p}}}
\newcommand{\Prec}{\ensuremath{P\rmsub{r}}}
\newcommand{\Tp}{\ensuremath{T\rmsub{p}}}
\newcommand{\gtildei}{\ensuremath{\tilde{g}_i}}
\newcommand{\epstildei}{\ensuremath{\tilde{\epsilon}_i}}
\newcommand{\taud}{\ensuremath{\tau\rmsub{d}}}
\newcommand{\taup}{\ensuremath{\tau\rmsub{p}}}
\newcommand{\tauc}{\ensuremath{\tau\rmsub{c}}}
\newcommand{\taum}{\ensuremath{\tau\rmsub{m}}}
\newcommand{\taua}{\ensuremath{\tau\rmsub{a}}}
\newcommand{\taun}{\ensuremath{T\rmsub{n}}}
\newcommand{\tauf}{\ensuremath{T\rmsub{f}}}

\newcommand{\sigmaR}{\ensuremath{\sigma\rmsub{R}}}
\newcommand{\sigmaa}{\ensuremath{\sigma\rmsub{a}}}
\newcommand{\sigmam}{\ensuremath{\sigma\rmsub{m}}}
\newcommand{\sigmaI}{\ensuremath{\sigma\rmsub{I}}}
\newcommand{\Deltamin}{\ensuremath{\Delta\rmsub{min}}}

\newcommand{\inquotes}[1]{\lq\lq #1\rq\rq}

\newcommand{\gatherindent}{\ \ \ \ \ \ }

\begin{document}

\title{Sustained sensorimotor control as intermittent decisions about prediction errors\thanks{This work was supported by Jaguar Land Rover and the UK-EPSRC grant EP/K014145/1 as part of the jointly funded Programme for Simulation Innovation (PSi), as well as UK-EPSRC grant EP/J002933/1 (FORWARN).}
}
\subtitle{Computational framework and application to ground vehicle steering}



\author{Gustav Markkula \and
        Erwin Boer \and
				Richard Romano \and
				Natasha Merat
}


\institute{G. Markkula \at
              Institute for Transport Studies, University of Leeds \\
              Tel.: +44 (0)113 34 39832\\
              \email{g.markkula@leeds.ac.uk}           
           \and
           E. Boer, R. Romano, and N. Merat \at
              Institute for Transport Studies, University of Leeds
}


\maketitle

\begin{abstract} 
A conceptual and computational framework is proposed for modelling of human sensorimotor control, and is exemplified for the sensorimotor task of steering a car. The framework emphasises control intermittency, and extends on existing models by suggesting that the nervous system implements intermittent control using a combination of (1) motor primitives, (2) prediction of sensory outcomes of motor actions, and (3) evidence accumulation of prediction errors. It is shown that approximate but useful sensory predictions in the intermittent control context can be constructed without detailed forward models, as a superposition of simple prediction primitives, resembling neurobiologically observed corollary discharges. The proposed mathematical framework allows straightforward extension to intermittent behaviour from existing one-dimensional continuous models in the linear control and ecological psychology traditions. Empirical observations from a driving simulator provide support for some of the framework assumptions: It is shown that human steering control, in routine lane-keeping and in a demanding near-limit task, is better described as a sequence of discrete stepwise steering adjustments, than as continuous control. Furthermore, the amplitudes of individual steering adjustments are well predicted by a compound visual cue signalling steering error, and even better so if also adjusting for predictions of how the same cue is affected by previous control. Finally, evidence accumulation is shown to explain observed covariability between inter-adjustment durations and adjustment amplitudes, seemingly better so than the type of threshold mechanisms that are typically assumed in existing models of intermittent control.  
\end{abstract}

\section*{List of recurring symbols}
\renewcommand{\arraystretch}{1.3}
\hspace{-.2cm}\begin{tabularx}{\columnwidth}{lX}
$A(t)$													& Evidence accumulator activity level \\
$A_+$; $A_-$										& Positive and negative decision thresholds for accumulator \\
$C(t)$													& Control generated by the human \\
$F(s)$													& Laplace domain transfer function from controlled system state to the perceptual quantity $P$ \\
$G(t)$													& Kinematic motor primitive \\
$g_i \triangleq K'\epsilon(t_i)$					& Expected (noise-free) amplitude of $i$th control adjustment \\
$\gtildei \triangleq K'\epstildei$				& Actual amplitude of $i$th control adjustment \\
$H(t)$													& Prediction primitive \\
$k$     												& Evidence accumulation input gain \\
$K$															& Gain in a generalised continuous control \\
$K' \triangleq$ $K \Delta T$ 		& Gain in a generalised intermittent control \\
\end{tabularx}

\hspace{-.6cm}\begin{tabularx}{\columnwidth}{lX}
$m_i$														& Motor noise affecting the $i$th control adjustment \\
$t_i$														& Onset time of the $i$th control adjustment \\
$\taun$; $\tauf$								& Preview time to near and far sight points (steering model) \\
$\knI$; $\knP$; $\kf$   				& Control gains for near point angle, near point rate, and far point rate (steering model) \\
$P(t)$													& Perceptual quantity describing the need for control (a negative control error) \\
$\Prec(t) \triangleq P(t - \taup)$ & Received $P(t)$, after perceptual delays \\
$\Pp(t)$												& Predicted value of $\Prec(t)$ \\
$Y(s)$													& Laplace-domain transfer function of the controlled system \\
$\delta(t)$											& Steering wheel angle applied by human (steering model) \\
$\Delta T$											& Control adjustment duration \\
$\Delta t_i \triangleq t_i - t_{i-1}$ & Time between two consecutive control adjustments \\
$\epsilon(t) \triangleq \Prec(t) - \Pp(t)$ & Control need prediction error (negative control error prediction error) \\
$\epstildei \triangleq (1 + m_i) \epsilon(t_i)$ & Prediction error at onset of $i$th control adjustment, after scaling with signal-dependent motor noise \\
$\thetan(t)$; $\thetaf(t)$	& Angles to near and far sight points in the vehicle's reference frame (steering model) \\
$\nu(t)$												& Evidence accumulation noise \\				
$\taup$ & Perception delay time \\	
$\taum$ & Motor delay time \\
$\tauc$ 																			& Control decision delay time in a generalised continuous model \\
$\taud \triangleq \taup + \tauc + \taum $			& Total control delay in a generalised continuous model \\						
$\sigmaa$ & Accumulator noise standard deviation \\
$\sigmam$ & Motor noise standard deviation \\
$\sigmaR$ & Road/vehicle noise standard deviation (steering simulations) \\
\end{tabularx}

\section{Introduction}
\label{sec:Introduction}

Many human sensorimotor activities that are sustained over time can be understood, on a high level, as the human attempting to control the body or the environment towards certain fixed or time-varying target states. Examples of such behaviours include postural control, tracking of external objects with eyes, hands or tools, and locomotion towards a target or along a path, either by foot or using some form of vehicle. In these types of behaviours, human behaviour has been likened to that of a servomechanism or controller \citep{Wiener1948}, and since the 1940s many mathematical models of human sensorimotor control behaviour have been proposed based on the continuous, linear feedback mechanisms of classical engineering control theory \citep[e.g.,][]{Tustin1947, McRuerEtAl1965, Nashner1972, RobinsonEtAl1986, KrauzlisAndLisberger1994, Peterka2000}. 

These basic ideas and models have been developed further in various directions. One line of investigation, building on notions from ecological psychology \citep{Gibson1986} or perceptual control theory \citep{Powers1978}, has investigated the nature of the exact information extracted by humans from their sensory input for purposes of control \citep[e.g.,][]{Lee1976, McBeathEtAl1995, SalvucciAndGray2004, Warren2006, ZagoEtAl2009, Marken2014}. An important goal in this field has been the identification of perceptual \emph{invariants}, which provide direct sensory access to task-relevant information \citep[e.g.,~the ratio between retinal size and expansion of an object is a good approximation of time to collision/interception;][]{Lee1976} and therefore lend themselves to simple but effective control \emph{heuristics}, typically formulated as one-dimensional linear control laws. 

Another important development has been the uptake of more modern control theoretic constructs, most notably \emph{optimal control theory} \citep{KleinmanEtAl1970, McRuer1980}. Optimal control models of sensorimotor behaviour suggest that humans act so as to minimise some cost function, typically weighing together control error and control effort, and theoretical predictions from these models have been confirmed experimentally \citep{TodorovAndJordan2002, LiuAndTodorov2007}. Typical engineering-inspired realisations of optimal control models include inverse and forward models of the controlled system \citep{ShadmehrAndKrakauer2008, FranklinAndWolpert2011}, but it remains contentious whether the nervous system has any such internal models, or whether it achieves its apparent optimality by means of other mechanisms \citep{Friston2011, PickeringAndClark2014}.

Another direction of research, which this paper aims to extend upon in particular, has emphasised the \emph{intermittency} of human control. Already early researchers noted that humans are not always continuously active in their sensorimotor control, but often instead seem to make use of intermittent, ballistic control adjustments \citep{Tustin1947, Craik1948}; \figref{TustinExample} provides an example. This mode of sensorimotor behaviour is well known from saccadic eye movements \citep[e.g., ][]{GirardAndBerthoz2005}, but has also been studied and evidenced in visuo-manual tracking \citep{MeyerEtAl1988, MiallEtAl1993, HannetonEtAl1997, PasalarEtAl2005, vandeKampEtAl2013}, inverted pendulum balancing \citep{LoramAndLakie2002, GawthropEtAl2013} and postural control \citep{CollinsAndDeLuca1993, LoramEtAl2005, AsaiEtAl2009}. A recurring suggestion in this work has been that control intermittency arises due to a minimum refractory time period that has to pass between consecutive bursts of control activity, and/or minimum control error thresholds that have to be surpassed before control is applied. 
Based on such assumptions, task-specific computational models of intermittent control have been proposed \cite[e.g.,][]{MeyerEtAl1988, CollinsAndDeLuca1993, MiallEtAl1993, BurdetAndMilner1998, GordonAndMagnuski2006, AsaiEtAl2009, MartinezGarciaEtAl2016}. 
However, the only complete, task-general, computational framework of intermittent control that we are aware of is that of Gawthrop and colleagues \citep{GawthropEtAl2011, GawthropEtAl2013, GawthropEtAl2015}. Their framework is an extension of the continuous optimal control theoretic models by \cite{KleinmanEtAl1970}, features forward and inverse models, and includes provisions allowing for both a minimum refractory period and error deadzones. 

\begin{figure}
\vspace{.3cm}
  \textbf{Reproduced figure not included in preprint; see Fig. 3 in \citep{Tustin1947}.}
\vspace{.3cm}
\caption{An early observation of intermittent-looking control by \cite{Tustin1947}. The plot is of the operator handle position in a gun turret aiming task. Note how a large fraction of the control signal plateaus with zero rate of change. } 
\label{fig:TustinExample}       
\end{figure}

This paper introduces an alternative computational framework for intermittent control, which was originally developed in the context of longitudinal and lateral control of ground vehicles. In that specific task context, the basic concepts have been described before \citep{Markkula2014, Markkula2015}. Here, the framework will be presented in a more general context, in the hope that it might prove useful also in other sensorimotor task domains. The framework ideas will also be developed for the first time in full mathematical detail, here, for the special case of one-dimensional control using stepwise control adjustments (further generalisation will be one topic in the Discussion). The main example will be an application of the computational framework to specify a model of car steering, and human steering data will be used for testing some of the framework's assumptions. 

The two main theoretical aims of this paper are: (1) To propose a framework for sustained, intermittent control that starts out from a classical control theory standpoint, without incorporating the extra assumptions typical of optimal control theory. 
This allows direct generalisation to intermittent control from existing psychological models based on perceptual invariants and control heuristics, and it also has some interest in light of the abovementioned debate about the neurobiological plausibility of optimal control theoretic models. 
(2) To propose a framework that actively connects with three concepts that are well established in contemporary neuroscience: \emph{motor primitives}, neuronal \emph{evidence accumulation}, and \emph{prediction of sensory consequences of motor actions}; these will all be introduced in further detail in the next section.
The use of any one of these three concepts in mathematical modelling of sensorimotor control is not novel in itself. However, to the best of our knowledge, the three have not previously been incorporated into one common framework. Such integration of  modelling concepts from different research fields (perceptual psychology, control theory, perceptual decision-making, motor control, etc.) necessarily involves some degree of simplification. Specialists in the fields we borrow from here will hopefully forgive component-level imperfections, in the interest of working towards a meaningful bigger picture.




\secref{ConceptualOverview} below will explain the three main concepts mentioned above, and briefly review to what extent they have been adopted in existing models of sensorimotor control, before \secref{ConceptualFramework} introduces the proposed framework on a conceptual, qualitative level. Then, \secref{Model} will present a computational realisation of the framework, for the special case of one-dimensional stepwise control, and briefly describe how it can be applied to a minimal example task, as well as to ground vehicle steering. Next, in \secref{ReconstructionMethod}, a simple signal reconstruction method will be described. This method, the proposed computational formulations, and two datasets of human steering of cars, will then be put to use in \secref{TestingModelPredictions}, providing some first empirical support for the framework. \secref{Discussion} will provide a discussion of the empirical and theoretical results, the relationship between the proposed framework and existing theories and models, as well as outline some possible future developments, before the conclusion in \secref{Conclusion}.

\section{Background}
\label{sec:ConceptualOverview}

\subsection{Motor primitives}

There is much emerging evidence for the idea that animal and human body movement is constructed from a fixed or only slowly changing repertoire of stereotyped pulses or synergies of muscle activation, which can be scaled in amplitude to the needs of the situation, and combined with each other, for example by linear superposition, to create complex body movement \citep{FlashAndHenis1991, FlashAndHochner2005, BizziEtAl2008, HartAndGiszter2010, Giszter2015}. Task-specific models have been proposed, where for example manual reaching \citep{MeyerEtAl1988, BurdetAndMilner1998} and car steering \citep{Benderius2014, MartinezGarciaEtAl2016} has been modelled as a sequence of superpositioned, ballistic motor primitives, for example bell-shaped pulses of movement speed. Furthermore, some authors have suggested task-general accounts describing motor control as constructed from such sequences of primitives \citep{HoganAndSternad2012, Karniel2013}. Here, we integrate this idea into a task-general, fully specified computational account.

It should be noted that the term \inquotes{motor primitive} has been used for a range of related but different concepts in the motor control literature; what we intend here could be further specified as \emph{kinematic} motor primitives, described by \cite{Giszter2015} as \inquotes{patterns of motion without regard to force or mass, e.g., strokes [...] or cycles [...]} (p.~156).

\subsection{Evidence accumulation}

\begin{figure}
  \includegraphics[width=\columnwidth]{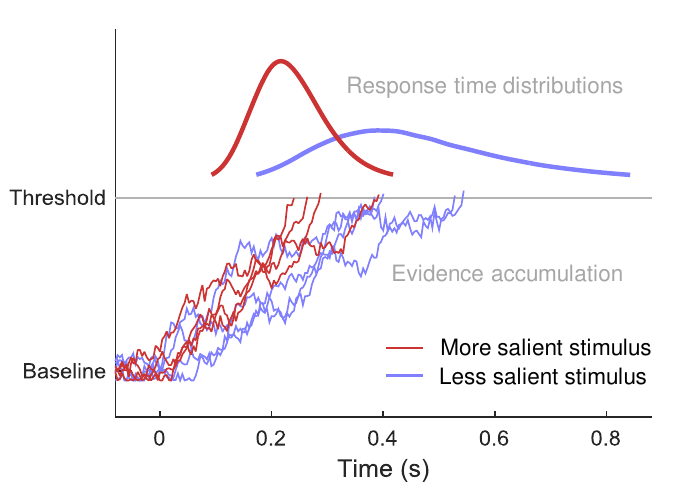}
\caption{A schematic illustration of how neuronal evidence accumulation mechanisms explain action onset timing distributions in perceptual decision-making tasks. After the onset of a stimulus ($t = 0$), noisy neuronal activity builds up over time. The reaching of a threshold activity level predicts overt action onset in individual trials, and stimulus saliency affects the rate of activity build-up. 
}
\label{fig:EvidenceAccumulationIllustration}       
\end{figure}

From laboratory paradigms on perceptual decision-making, where humans or animals have to interpret sensory input to decide on a single correct motor action, there is strong behavioural and neuroimaging evidence suggesting that the initiation of the motor action occurs when neuronal firing activity in task-specific neurons has accumulated to reach a threshold, with noise in the accumulation process explaining action timing variability \citep{Ratcliff1978, UsherAndMcClelland2001, CookAndMaunsell2002, GoldAndShadlen2007, PurcellEtAl2010}; see \figref{EvidenceAccumulationIllustration} for an illustration. Importantly, the more unambiguous and salient the stimulus being responded to, the quicker the rate of increase of neuronal activity \citep[e.g., ][]{Ditterich2006, PurcellEtAl2010, PurcellEtAl2012}. It has been shown that by properly adapting the parameters of such an evidence accumulation to the task at hand, including sensory noise levels, the brain could use this type of mechanism to achieve Bayes-optimal perceptual decision-making \citep{BogaczEtAl2006, BitzerEtAl2014}. 

A novel contribution of the present framework is the suggestion, conceptually and computationally, that (1) sustained sensorimotor control can be regarded as a sequence of such perceptual-motor decisions, and (2) the magnitude of control errors (among other things) might affect the rate of  evidence accumulation. These suggestions are in contrast with existing models of intermittent control, which, as mentioned above, assume that control adjustment timing is determined by thresholds on control errors and/or inter-adjustment time durations.

\subsection{Prediction of sensory outcomes of motor actions}
\label{sec:PredictionConcept}

It has been shown in both primates and other animals that whenever a movement command is issued in the nervous system, it tends to be accompanied by a so-called \emph{corollary discharge} (possibly mediated by an \emph{efference copy} of the movement command), biasing sensory brain areas whose inputs will be affected by the motor action in question. There is much evidence to support the idea that these biases are predictions of sensory consequences of the motor action, which might allow the nervous system to infer whether incoming sensory stimulation is due to the organism's own actions or to external events \citep{Sperry1950, VonHolstAndMittelstaedt1950, PouletAndHedwig2007, CrapseAndSommer2008AnimalKingdom, AzimAndAlstermark2015}. For example, the image of the outside world translating over the retina could mean either that the outside world is rotating, or that that the eyes are. 

In sensorimotor control, a specific use of such a discriminating function could be to deal with time delays in the control loop, in a manner similar to the Smith Predictor in engineering control theory \citep{Smith1957, MiallEtAl1993}: After initiating a control action to address a control error, the correct prediction for a time-delayed system is that the error will not disappear immediately, and as long as the control error responds as predicted over time, there is no need to infer that the situation in the external world has changed to warrant further control action than what has already been applied. Such a mechanism
would seem to be particularly beneficial for intermittent control, but so far seems to have received less modelling attention in this type of context \citep{BurdetAndMilner1998} than in continuous control \citep[e.g., ][]{KettnerEtAl1997, ShadmehrAndKrakauer2008, FristonEtAl2010, GrossbergEtAl2012}. 

Here, besides integrating the familiar Smith Predictor mechanism in the computational framework, it will also be shown how a prediction signal that is useful in the intermittent control context can be generated similarly to the intermittent control  itself, as a superposition of simple primitives. Neuronal recordings from animals show time histories of corollary discharge biases that follow a general pattern of rapid initial increase followed by slower decay \citep{PouletAndHedwig2007, ChagnaudAndBass2013, RequarthAndSawtell2014}; interestingly the near-optimal \inquotes{prediction primitives} proposed here for intermittent sensorimotor control take a similar form.

\section{A conceptual framework for intermittent control}
\label{sec:ConceptualFramework}

On a conceptual level, what is being proposed here is that sustained sensorimotor control can be understood and modelled as a combination of the three mechanisms described above, as follows: Perceptual cues (e.g., invariants) that indicate a need for control--i.e., which indicate control error--are considered in a decision-making process that can be modelled as noisy accumulation towards a threshold. At this threshold, a new control action is initiated, in the form of a ballistic motor primitive that is superpositioned, linearly or otherwise, onto any other ongoing motor primitives. The exact motor primitive that is initiated is the one that the nervous system has reason to believe will be most appropriate, based on the available perceptual data and previous experiences. An important part of selecting an appropriate motor primitive might be a heuristic scaling of the primitive's amplitude with the magnitude of the perceived control error. At motor primitive initiation, a prediction is also made, for example in the form of a corollary discharge, of how the control error will be reduced over time thanks to the new control action. This new prediction is superpositioned onto any previously triggered predictions. The resulting overall prediction signal inhibits (is subtracted from) the control error input, such that what the intermittent control is reacting to (what is being accumulated; what the control actions are scaled by) is actually \inquotes{control error prediction error} rather than control error per se. 

The next section develops this conceptual account into a computational one, for the special case of one-dimensional control using stepwise adjustments of a stereotyped shape, and also shows how it relates to more conventional, continuous linear control models.

\section{Computational framework for stepwise one-dimensional control}
\label{sec:Model}

\subsection{Task-general formulation}
\label{sec:TaskGeneralModel}

\begin{figure*}
  \includegraphics[width=\textwidth]{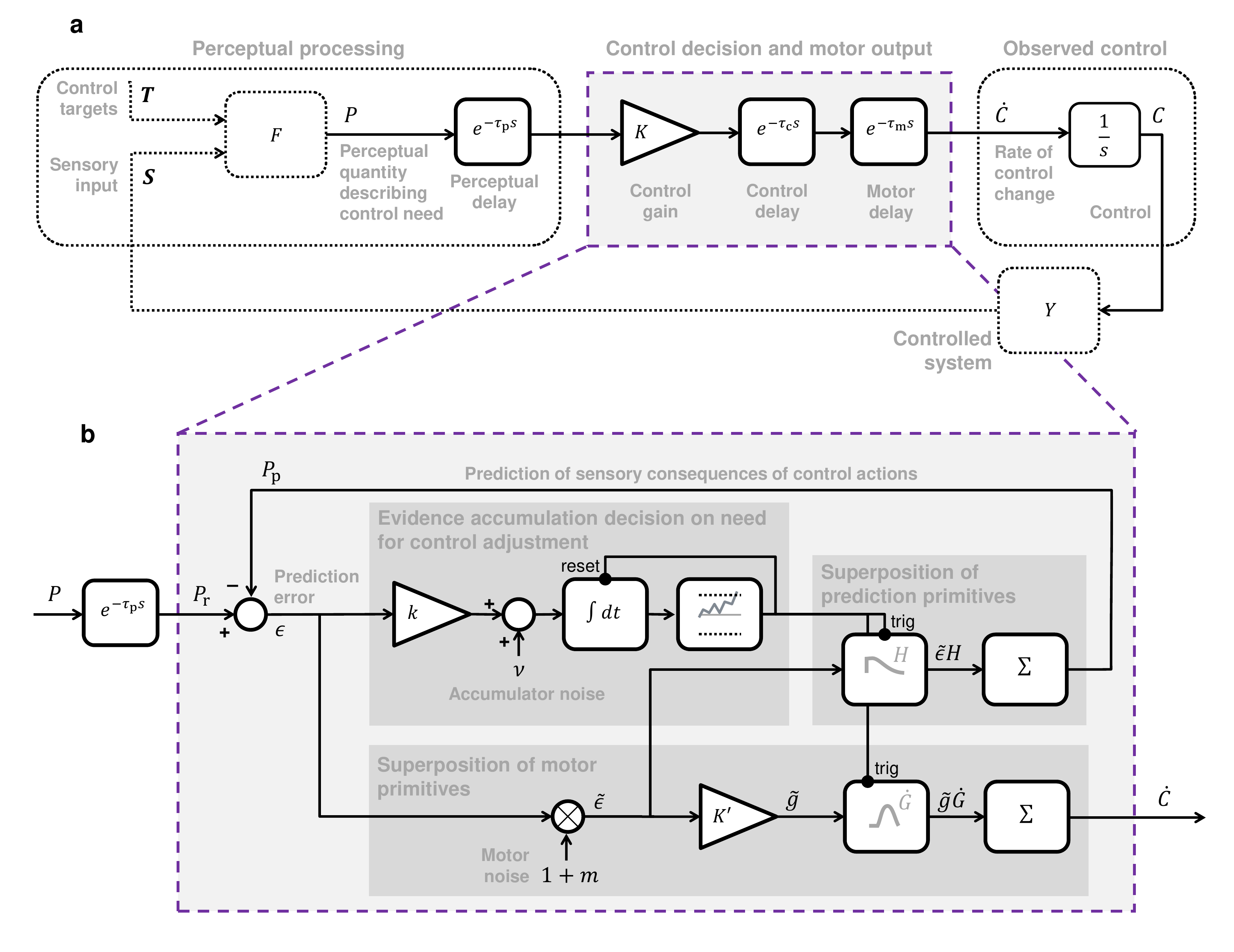}
	\caption{An illustration of how a continuous, linear control law (panel a) can be generalised to intermittent control using the computational framework proposed here (panel b).}
	\label{fig:Flowcharts}       
\end{figure*}

A very general formulation of \emph{continuous} one-dimensional sensorimotor control is sketched in \figref{Flowcharts}(a). The human is assumed to process sensory inputs $\mathbf{S}(t)$ and control targets $\mathbf{T}(t)$ over time $t$, to yield a one-dimensional quantity $P(t)$, that when delayed and multiplied by a gain $K$, yields the rate of change $\dot{C}(t)$ of the control to be applied:
\begin{equation}
\label{eq:GeneralisedContinuousControl}
\dot{C}(t) = K \cdot P(t - \taud),
\end{equation} 
where $\taud \triangleq \taup + \tauc + \taum$ is a sum of delays at perceptual, control decision, and motor stages, and where a positive $\dot{C}$ changes the control in a direction that tends to change $P$ in a negative direction, and vice versa. The control thus strives to reduce $P$ to zero, such that $P$ can be construed as a perceptual invariant quantifying a negative \emph{control error}, or, differently put, quantifying the \emph{need for a change in control}. Typically, this quantification will be non-exact and heuristic. Note that the control gain $K$ can just as well be absorbed into the $P$ function by fixing $K = 1$ above, which gives $P$ an even more specific interpretation as the \emph{needed rate of control change} in the given situation, in units of $\dot{C}$. Among these various interpretations of $P$ we will mainly refer to it as a \inquotes{perceptual control error}, for ease of reading, and to emphasise the connection to classical control theory.

Note also that control laws that are mathematically equivalent to \eqnref{GeneralisedContinuousControl} can be obtained by differentiation or integration with respect to time, to instead model control in terms of for example $C$ or $\ddot{C}$.

As suggested in \figref{Flowcharts}(b), the computational framework being proposed here can be understood as replacing the \inquotes{control decision and motor output} component of this type of continuous model with the mechanisms that were outlined in \secref{ConceptualOverview}, to generate control that is intermittent, but which in many circumstances will be rather similar in appearance to the continuous control \citep[cf.][]{GawthropEtAl2011}.


Below, the different parts of the framework will be defined in detail.

\subsubsection{Perceptual control error quantity}

What is being proposed here is independent of what specific quantity $P$ might appropriately quantify the human's perceived need for control in the task at hand. 
In contrast, in many continuous models of human control, the main modelling challenge has in practice been to define a $P$ such that \eqnref{GeneralisedContinuousControl} reproduces observed human behaviour as closely as possible. 
Below, some examples of continuous models from the literature will be provided, all of which can be written on the form of \eqnref{GeneralisedContinuousControl}, thus making them all candidates for generalisation from continuous to intermittent control as proposed here.

For example, for a task of manually tracking a one-dimensionally moving target with a mouse cursor, \cite{Powers2008} showed that the rate of mouse cursor movement could be well described as proportional to the distance $D(t) =  C(t) - C\rmsub{T}(t)$ between actual and target cursor position: 
\begin{equation}
\label{eq:MouseTrackingControl}
\dot{C}(t) = -K \cdot D(t - \taud),
\end{equation}
i.e.~in this case we get simply $P(t) = -D(t)$. 

A more general example can be had from McRuer and colleagues \citep{McRuerEtAl1965, McRuerAndJex1967}, who, based on their work on the so-called \emph{cross-over} model, suggested the following generalised Laplace domain expression for a human controller responding to a control error $e$:
\begin{equation}
C(s) = -K \frac{T_L s + 1}{T_I s + 1} e^{-\Delta T s} e(s).
\end{equation}
where $T_L$ and $T_I$ are lead and lag time constants. Rewriting to time domain:
\begin{equation}
T_I \dot{C}(t)  + C(t) = -K \left[ T_L \dot{e}(t - \taud) + e(t - \taud) \right], 
\end{equation}
we see that in this case we can write:
\begin{equation}
P(t) = \frac{-K\left[T_L\dot{e}(t) + e(t)\right] - C(t + \taud)}{K T_I}.
\end{equation}
Note that in this expression, the rate of control change that will be applied, after the total neuromuscular delay $\taud$, depends also on the control value, and more precisely on what the control value will be just before the new control rate comes into effect.

Another example of this type of rewriting of continuous controllers to the form of \eqnref{GeneralisedContinuousControl} is the PID-controller type model of upright postural control (quiet standing) proposed by \citeauthor{Peterka2000} \citep{Peterka2000, MaurerAndPeterka2005}:
\begin{equation}
P(t) = -\left( K\rmsub{I}\theta(t) + K\rmsub{P}\dot{\theta}(t) + K\rmsub{D}\ddot{\theta}(t) \right),
\end{equation}
with $\dot{C}$ now the rate of change of a balancing torque around the ankle joint, and where $\theta$ is the body sway angle. Yet another example is the ecological psychology-based vehicle steering model by \cite{SalvucciAndGray2004}:
\begin{equation}
\label{eq:SalvucciAndGray}
P(t) = \knI \thetan(t) + \knP \thetandot(t) + \kf \thetafdot(t),
\end{equation}
with $\dot{C}$ being rate of steering wheel angle change, and where $\thetan$ and $\thetaf$ are visual angles to two reference points on the road, one \inquotes{near} and one \inquotes{far}. Note that in both of these latter two models, there are control gain parameters (the $K\rmsub{\bullet}$ and $k\rmsub{\bullet}$) for all of the terms in $P(t)$, so one can fix $K = 1$ in \eqnref{GeneralisedContinuousControl}, as mentioned above.

\subsubsection{Evidence accumulation}

When to perform a control adjustment is modelled here as a process of \emph{two-sided} evidence accumulation \citep[or \emph{drift diffusion;}][]{Ratcliff1978, RatcliffAndMcKoon2008}. In this type of model, the accumulation of strictly positive neural firing rates, as schematically illustrated in \figref{EvidenceAccumulationIllustration}, is replaced by accumulation of a quantity that can be either positive or negative, with one threshold on either side of zero, $A_+$ and $A_-$, representing two different alternative decisions \citep[this is mathematically equivalent to for example having two mutually inhibitory one-sided accumulators;][]{BogaczEtAl2006}. In the present context of one-dimensional control, these two thresholds represent decisions to make a control adjustment in either of the two possible directions of control. Such an accumulator can be defined in many different ways. One rather general possible formulation, based on \citep{BogaczEtAl2006} and \citep{PurcellEtAl2010}, would be:
\begin{equation}
\label{eq:GeneralisedAccumulator}
\frac{\mathrm{d}A(t)}{\mathrm{d}t} = \gamma \left[ \eta \left( \epsilon(t) \right) \right] - \lambda A(t) + \nu(t), 
\end{equation}
where $A(t)$ is the activation of the accumulator, $-\lambda A(t)$ is a leakage term, and $\nu(t)$ is noise, for example Gaussian white noise with zero mean and variance $\sigmaa^2 \Delta t$ across a simulation time step of duration $\Delta t$. Furthermore, $\epsilon$ is the error in predicted control error:
\begin{equation}
\label{eq:epsilon}
\epsilon(t) \triangleq \Prec(t) - \Pp(t),
\end{equation}
where $\Pp(t)$ is the brain's prediction, to be defined in detail in \secref{ControlErrorPrediction}, of the perception-delayed control error quantity $\Prec(t) \triangleq P(t - \taup)$. 
Finally, $\eta(\epsilon)$ in \eqnref{GeneralisedAccumulator} is an activation function, for example sigmoidal, and $\gamma$ is a gating function, zero for small input values, for example defined as:
\begin{equation}
\gamma(\eta) = \mathrm{sgn}(\eta) \cdot \max(0, |\eta| - \eta_0)
\end{equation}
In the example implementations of the framework proposed further below, the accumulators are simplified special cases of \eqnref{GeneralisedAccumulator} with $\eta_0 = \lambda = 0$, i.e.~without gating or leakage, and with $\eta(\epsilon) = k \epsilon$, where $k$ is a gain parameter, thus reducing the accumulation equation to
\begin{equation}
\label{eq:SimplifiedAccumulator}
\frac{\mathrm{d}A(t)}{\mathrm{d}t} = k \epsilon(t) + \nu(t), 
\end{equation}
which is also what is illustrated in \figref{Flowcharts}. 

As for the thresholds of the accumulator, it will in most control tasks probably make sense to select these to be of equal magnitude ($|A_+| = |A_-|$), and if so then these can both be set to unity magnitude without loss of generality ($A_+ = 1; A_- = -1$), since the accumulator activation is specified in arbitrary units.

\subsubsection{Control adjustments}
\label{sec:ControlAdjustments}

Upon reaching one of the accumulator thresholds, the accumulator is assumed to be reset to zero, and a new control adjustment primitive is generated (the \inquotes{reset} and \inquotes{trig} connections in \figref{Flowcharts}(b)). In the framework formulation being proposed here, all control adjustments have the same general shape $G$, which could be any function which starts out at zero and, after an initial motor delay $\taum$, rises to unity over the adjustment duration of $\Delta T$, i.e. any function which fulfils:  
\begin{equation}
\label{eq:G}
	G(t) = 
	\begin{cases}
		0  & \text{for } t \leq \taum \\
		1  & \text{for } t \geq \taum + \Delta T
	\end{cases}
\end{equation}
Consequently, the rate of change of control during a control adjustment is given by a function $\dot{G}$ that fulfils:
\begin{equation}
\label{eq:Gdot}
	\begin{cases}
		\dot{G}(t) = 0  \text{ for } t \leq \taum \text{ and } t \geq \taum + \Delta T\\
		\int_{\taum}^{\taum + \Delta T}{\dot{G}(t) \mathrm{d}t} = 1
	\end{cases}
\end{equation}
For example, as hinted at in \figref{Flowcharts}(b), $\dot{G}(t)$ could be a bell-shaped pulse beginning after a motor delay ($\taum$).

The expected value of the amplitude for the $i$th adjustment, beginning at the time $t_i$ at which the accumulator threshold was exceeded for the $i$th time, is obtained as:
\begin{equation}
\label{eq:IdealAdjustmentMagnitude}
g_i \triangleq K \Delta T \epsilon(t_i) \triangleq K' \epsilon(t_i) = K' \left( \Prec(t_i) - \Pp(t_i) \right).
\end{equation}
The relationship introduced above,
\begin{equation}
\label{eq:KAndKPrimeRelationship}
K \triangleq K' / \Delta T,
\end{equation}
between the gains of the continuous and intermittent controls, is not a crucial part of the model as such, but ensures that the two controls will typically be close approximations of each other. To see this, consider that for $\Pp \approx 0$, $\epsilon \approx \Prec \approx P$, such that the intermittent control will respond to a control error $P$ by adjusting the control by approximately $K' P$ in a time duration $\Delta T$, i.e., with an \emph{average} rate of change of control of
\begin{equation}
K'P / \Delta T = K P,
\end{equation}
which is also the control rate being applied by the continuous model around the same point in time.

In relation to the earlier discussion about the meaning of $P$ when absorbing all control gains into it, note that fixing $K' = 1$ in \eqnref{IdealAdjustmentMagnitude} makes $P$ a quantification of \emph{needed control adjustment amplitude}, in units of $C$.

Adding to \eqnref{IdealAdjustmentMagnitude} also motor noise, for example of a signal-dependent nature, whereby larger control movements will be more likely to have large inaccuracies \citep{FranklinAndWolpert2011}, one can write the actual control adjustment amplitude:
\begin{equation}
\gtildei \triangleq K' \epstildei,
\end{equation}
where:
\begin{equation}
\label{eq:epsilontilde}
\epstildei \triangleq (1 + m_i) \epsilon(t_i),
\end{equation}
with $m_i$ drawn from a normal distribution of zero mean and variance $\sigma_m^2$. 

Each new control adjustment is linearly superpositioned onto any adjustments that might be ongoing since previously \citep[see e.g.][]{FlashAndHenis1991, HoganAndSternad2012, Karniel2013, Giszter2015}, yielding an output rate of control:
\begin{equation}
\label{eq:ControlRateSummation}
\dot{C}(t) = \sum_{i=1}^n \gtildei \dot{G}(t - t_i),
\end{equation}
and therefore:
\begin{equation}
\label{eq:ControlSummation}
C(t) = C_0 + \sum_{i=1}^n \gtildei G(t - t_i),
\end{equation}
where $n$ is the total number of adjustments generated so far, and $C_0$ is an initial value of the control signal.

\subsubsection{Prediction of control error}
\label{sec:ControlErrorPrediction}

The prediction $\Pp(t)$ of the perceptual control error quantity $P(t)$ is generated by a similar superposition:
\begin{equation}
\label{eq:Pp}
\Pp(t) = \sum_{i=1}^n \epstildei H(t - t_i)
\end{equation}
where $H(t)$ is a function describing how, in the human's experience, control errors typically become corrected over time by a control adjustment, in the task at hand\footnote{One might also consider using the non-noisy $\epsilon_i$ in \eqnref{Pp}. We have chosen to use $\epstildei$ mainly because we have no other direct representation in the framework of sensory noise affecting control amplitudes and predictions. In this sense \eqnref{epsilontilde} and $m_i$ model both sensory and motor noise; improved frameworks should tease these things apart.}. By analogy with \eqnref{ControlSummation}, $H$ could be termed a \emph{prediction primitive}, and it is proposed here that this function should satisfy:
\begin{equation}
\label{eq:HRequirements}
	\begin{cases}
	H(t) = 0 & \text{for } t \leq 0 \\
	H(t) \rightarrow 1 & \text{for } t \rightarrow 0^+ \\
	H(t) = 0  & \text{for } t \geq \Tp,
	\end{cases}
\end{equation}
where $\Tp$ is the typical time from the triggering of a control adjustment until the controller receives evidence that the control error in question might have become completely corrected. The $t>0$ part of the $H$ function should describe how the perceptual control error quantity is expected to respond over time to the control adjustment. Mathematically, this part of $H$ should thus be something like the following:
\begin{equation}
\label{eq:TheoreticalH}
H(s) = 1 - G(s)Y(s)F(s)e^{-\taup s},
\end{equation}
It is however not necessary to assume that the brain calculates something like \eqnref{TheoreticalH} in detail. In practice, it might suffice to have a rather approximate $H$, for example describing a sigmoidal fall to zero, such as hinted at in \figref{Flowcharts}(b). 

There are two further specific assumptions motivating the exact formulations of \eqnsref{Pp} and \eqnrefjustpar{HRequirements}. First, immediately after the $n$th control adjustment has been initiated at time $t_n$, in absence of motor noise it is assumed that the predicted control error should be equal to the actual current control error, i.e.:
\begin{equation}
\label{eq:PpEqualsPAfterAdjustment}
\lim_{\Delta t \rightarrow 0^+} \Pp(t_n + \Delta t) = P(t_n), 
\end{equation}
i.e.~after a new adjustment, the prediction should \inquotes{acknowledge}, and start from, the currently observed control error. Second, over time, predicted control error should fall to zero. That the latter holds true with the proposed formulations is easy to see; it is a trivial consequence of requiring $H(t > \Tp) = 0$. To see that the former assumption is realised, one can write:
\begin{gather}
\lim_{\Delta t \rightarrow 0^+} \Pp(t_n + \Delta t) = \left\{ \mathrm{\eqnref{Pp}} \right\} \nonumber \\
\ \ \ \ \ \  = \lim_{\Delta t \rightarrow 0^+} \sum_{i=1}^n \epstildei H(t_n + \Delta t - t_i) \\
\ \ \ \ \ \  = \lim_{\Delta t \rightarrow 0^+}\sum_{i=1}^{n-1} \epstildei H(t_n + \Delta t - t_i) + \tilde{\epsilon}_n H(\Delta t) \\
\ \ \ \ \ \  = \left\{ \mathrm{\eqnsref{Pp} \ and \ \eqnrefjustpar{HRequirements}} \right\} \nonumber \\
\ \ \ \ \ \  = \Pp(t_n) + \tilde{\epsilon}_n \\
\ \ \ \ \ \  = \left\{ \mathrm{\eqnsref{epsilon} \ and \ \eqnrefjustpar{epsilontilde}} \right\} \nonumber \\
\ \ \ \ \ \  = \Pp(t_n) + (1 + m_i)\left[ P(t_n) - \Pp(t_n) \right] \label{eq:PpWithMotorNoise} \\
\ \ \ \ \ \  = P(t_n), \ \mathrm{for} \ m_i = 0
\end{gather}

It should be noted that, if the prediction $H$ is exact, the linear superpositions in \eqnsref{ControlRateSummation} through \eqnrefjustpar{Pp} should provide (near-)exact overall predictions for controlled systems that are (near-)linear (i.e.~for which a superposition of several individual control adjustments yields a system response which is exactly or approximately a superposition of how the system would have responded to each control adjustment separately).

In the next two subsections, the computational framework introduced above will be further explained and illustrated by means of two task-specific implementations.

\subsection{A minimal example}
\label{sec:MinimalExample}

\begin{figure*}
\subfloat[Step target]{
  \includegraphics[width=0.5\textwidth]{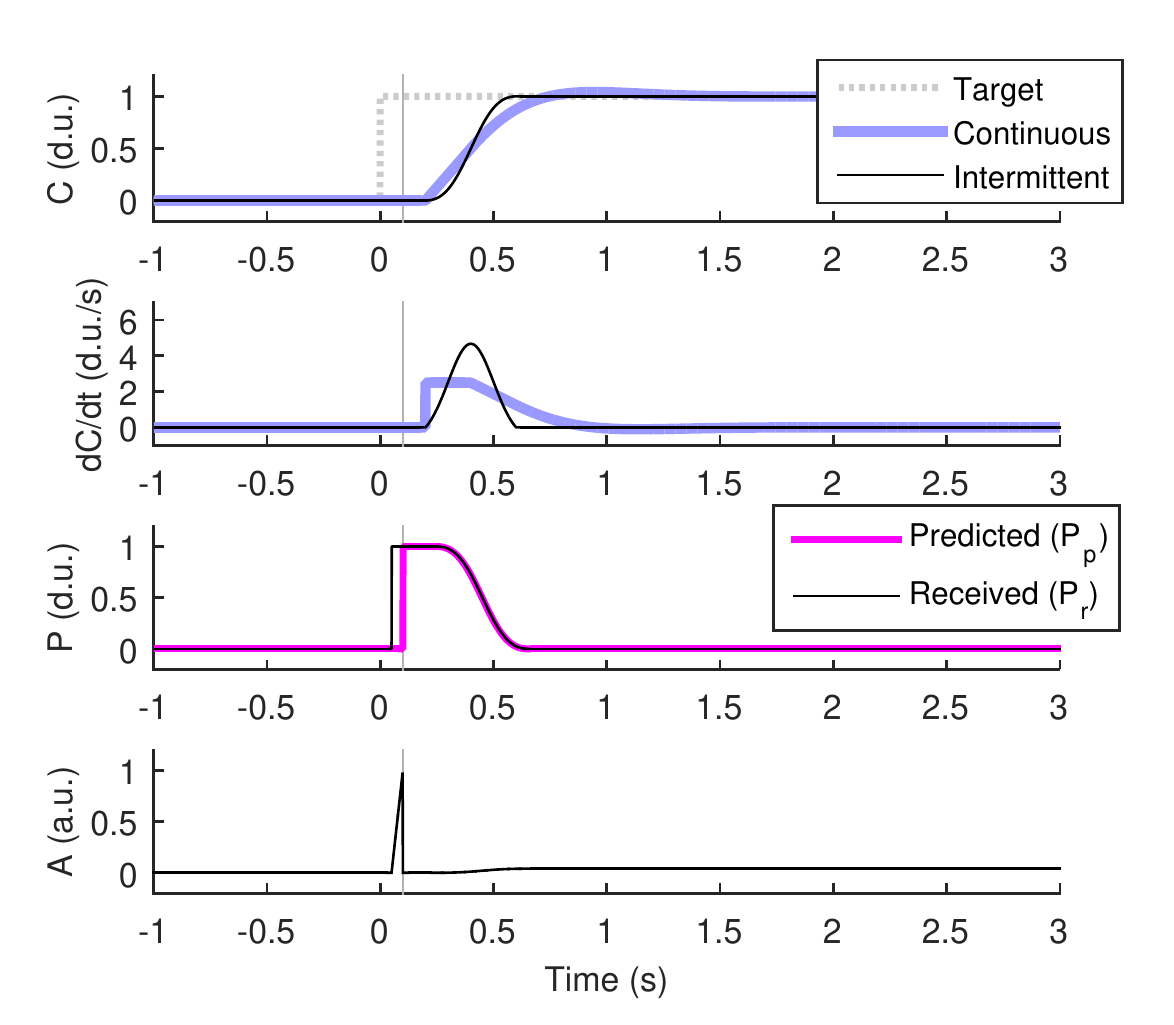}
}
\subfloat[Sum-of-sines target]{
  \includegraphics[width=0.5\textwidth]{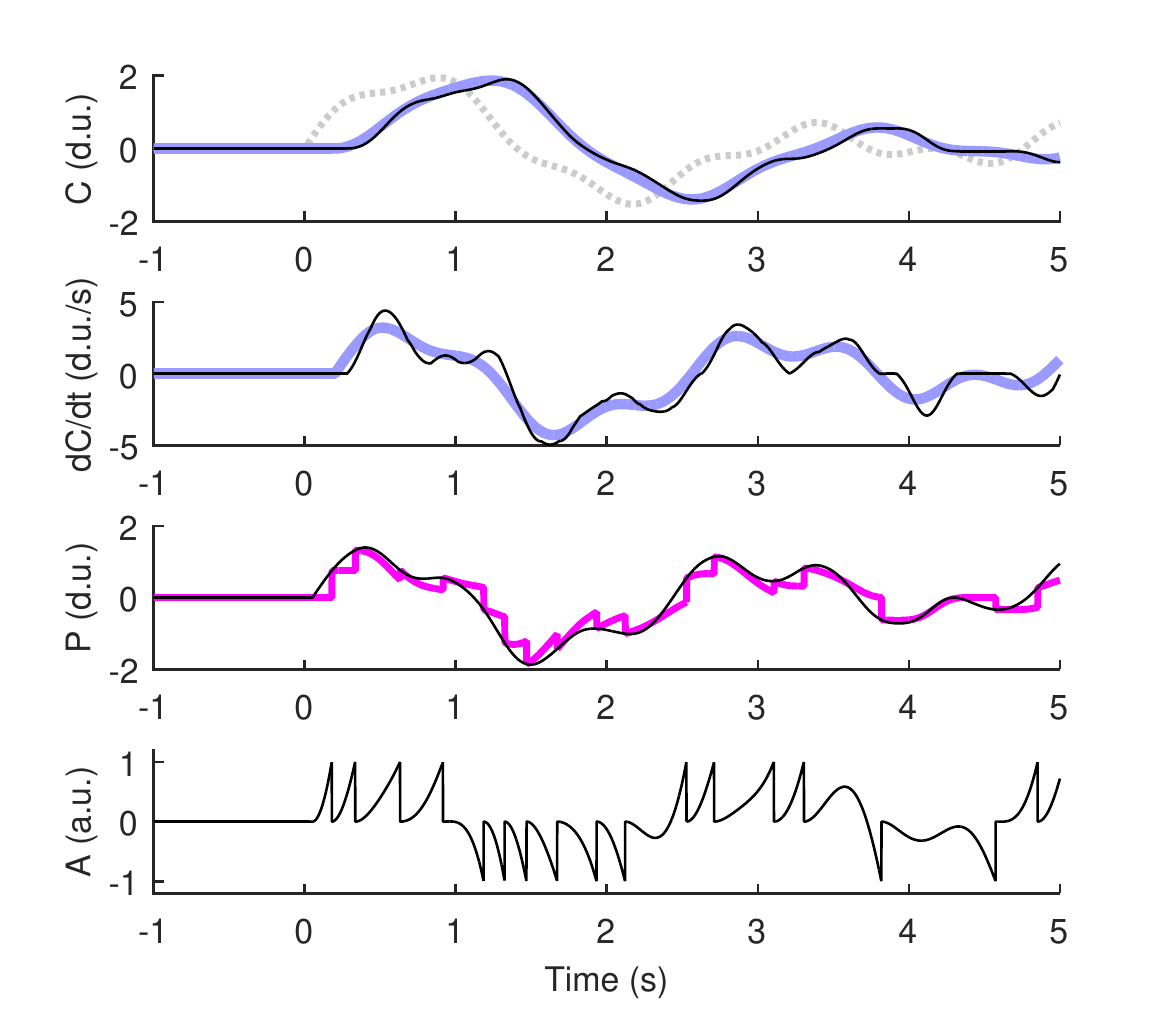}
}
\caption{Simulations of a continuous model by \cite{Powers2008}, of tracking an on-screen cursor with a mouse, as well as a generalisation of the same model to intermittent control, using the computational framework proposed here. In these examples, the intermittent control model is simulated completely without noise. In panel (a), the grey vertical line shows the time ($t$ = 0.1 s) at which the evidence accumulator (bottom panel) reaches threshold. 
}
\label{fig:MinimalExample}       
\end{figure*}

Consider the simple continuous control model by \cite{Powers2008} in \eqnref{MouseTrackingControl}, of a human tracking a target on a screen with a mouse cursor. The panels of \figref{MinimalExample} show, in light blue, the response of this model, with $K' = 1$, $\Delta T = 0.4$ s (making $K = K' / \Delta T = 2.5$), and $\taud = 0.2$ s, to a step input (panel a) and a more complex \inquotes{sum-of-sines} input (panel b). 

\begin{figure}
	\includegraphics[width=\columnwidth]{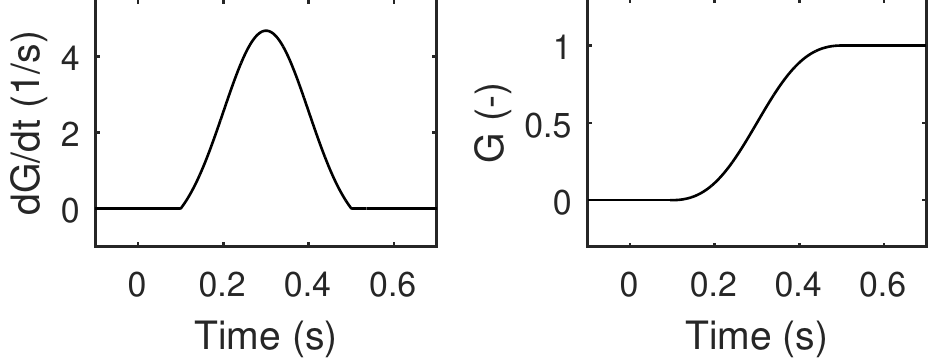}
	\caption{The bell-shaped control adjustment profile used for both the minimal cursor-tracking example in \secref{MinimalExample} and for the ground vehicle steering model.}
	\label{fig:GAndGdot}
\end{figure}

Also shown in \figref{MinimalExample}, in black, is the behaviour of the same model when generalised to intermittent control, using the computational framework described above. Here, perceptual and motor delays were set to $\taup = 0.05$ s and $\taum = 0.1$ s, based on \citep{LamarreEtAl1981, CookAndMaunsell2002, MorrowAndMiller2003, PurcellEtAl2010}\footnote{In more detail: The results by both \cite{CookAndMaunsell2002} and \cite{PurcellEtAl2010} suggest that decision-making neuronal activity accumulation begins about 50-100 ms after a visual stimulus onset. \cite{LamarreEtAl1981} observed an average lag between earliest primary motor cortex activity onset \citep[which might precede accumulator threshold-reaching somewhat;][]{SelenEtAl2012} and elbow movement onset of 122 ms, and \cite{MorrowAndMiller2003} observed an average lag of 50 ms between primary motor cortex activity and arm EMG signals (thus not taking into account any delays between threshold-reaching and movement-generating primary motor cortex activity, and between EMG and limb movement). In practice, the sum $\taup+\taum$ matters for model control behaviour, but not the individual terms.}, the accumulator gain was $k = 20$, the accumulator thresholds were at positive and negative unity, and all the other parameters of the accumulator were set to zero (i.e. no gating, leakage, or noise). As shown in \figref{GAndGdot}$,\dot{G}$  was, after the initial $\taum$ delay, $\pm 2$ standard deviations of a Gaussian, making $G$ reminiscent of (although not identical to) a minimum-jerk movement \citep{Hogan1984}. As for $H$, since in this task the control signal $C$ is also the quantity being controlled (with appropriate units for mouse and cursor position, and disregarding any delays between them), $Y(s)F(s) = 1$, and \eqnref{TheoreticalH} suggests the following error prediction function:
\begin{equation}
H(t) = \begin{cases}
	0, & t \leq 0 \\
	1 - G(t - \taup), & t > 0.
	\end{cases}
\end{equation}
As can be seen in the third panel of \figref{MinimalExample}(a), $H$ here thus specifies that after a control adjustment has been applied, $\Pp$ is first just set to $\Prec$, acknowledging the control error, then stays at this level for a period $\taum$, before the control adjustment begins, and then an additional $\taup$, while the effects of the adjustment feed through the perceptual system. Thereafter, $\Pp$ simply follows the shape of $G$ down to a zero predicted error. It may be noted that this shape of $H$ bears some resemblance to typical time courses of corollary discharge inhibition, as discussed in \secref{PredictionConcept}.

In \figref{MinimalExample}(a), note that the onset of control is equally delayed for the continuous and intermittent controllers, due to the parameter values for $k$ and $A_+$ being such that a unity control error accumulates to threshold in a time $\taua = 1/k = 0.05$ s, i.e.~$\taup + \taua + \taum = 0.2$ s, the same as the $\taud$ for the continuous model. In \figref{MinimalExample}(b), note that control adjustments often partially overlap, in linear superposition, to yield a less obviously stepwise resulting signal. Furthermore, note that the rate of control $\dot{C}$ for the continuous model indeed looks much like an average-filtered version of the $\dot{C}$ for the intermittent model (as discussed in \secref{ControlAdjustments}). Therefore, if a human would behave as the intermittent controller, the continuous model would still fit the observed behaviour very well. In the terms of \cite{GawthropEtAl2011}, the intermittent control \inquotes{masquerades} well as the continuous control. As discussed by \cite{Benderius2014}, such an underlying control intermittency might potentially be able to account for much of the nonlinear \inquotes{remnant} that is left unexplained by the continuous model.

\subsection{Application to ground vehicle steering}´
\label{sec:ApplicationToSteering}
For the specific sensorimotor task of steering a car, research and control model development have followed the same general directions outlined in the Introduction, with examples of both classical control theoretic models \citep{McRuerEtAl1977, Donges1978, Jurgensohn2007}, ecological psychology models \citep{FajenAndWarren2003, WannAndWilkie2004, WilkieEtAl2008}, optimal control models \citep{MacAdam1981, SharpEtAl2000, PlochlAndEdelmann2007}, and more recently also intermittent control models \citep{GordonAndMagnuski2006, RoyEtAl2008, Benderius2014, Markkula2014, GordonAndSrinivasan2014, GordonAndZhang2015, JohnsAndCole2015, BoerEtAl2016, MartinezGarciaEtAl2016}. 

\begin{figure*}
	\subfloat[Response of noise-free model to a 1\degree{} initial heading error.]{
		\includegraphics[width=0.33\textwidth]{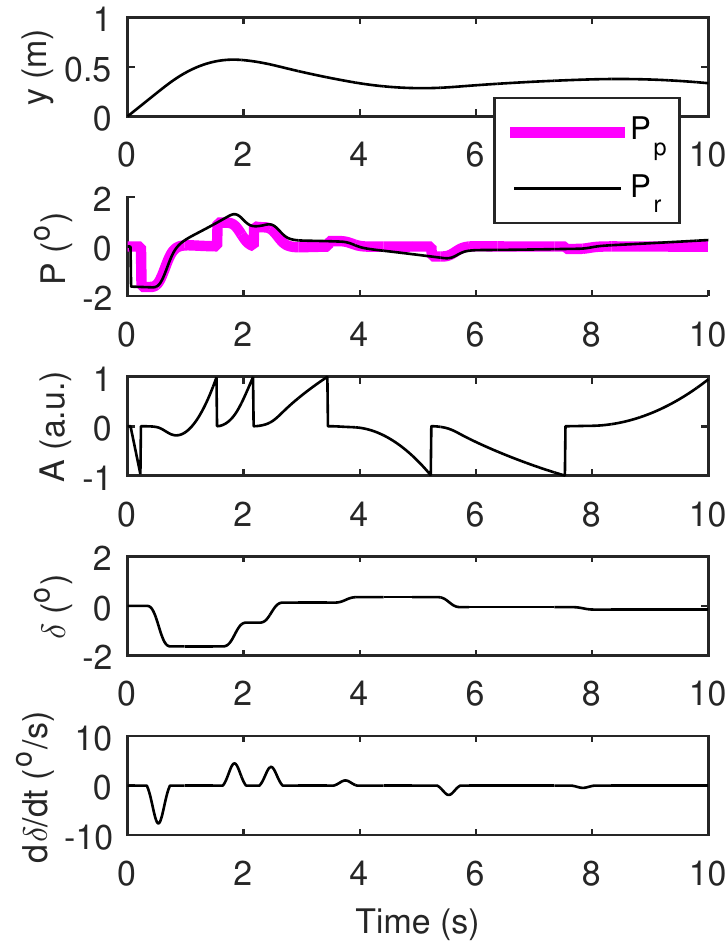}
	}
	\hspace{0.1cm}
	\subfloat[As in (a), but also including accumulator and motor noise.]{
		\includegraphics[width=0.313\textwidth]{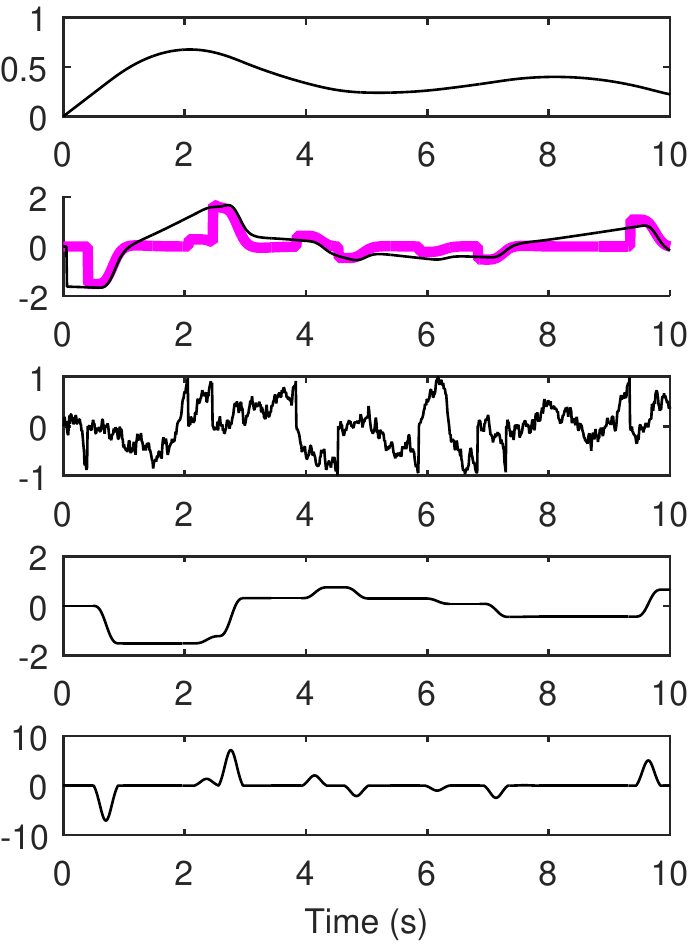}
	}
	\hspace{0.1cm}
	\subfloat[A 10 s excerpt of a longer lane-keeping simulation, with accumulator, motor, and road/vehicle noise.]{
		\includegraphics[width=0.313\textwidth]{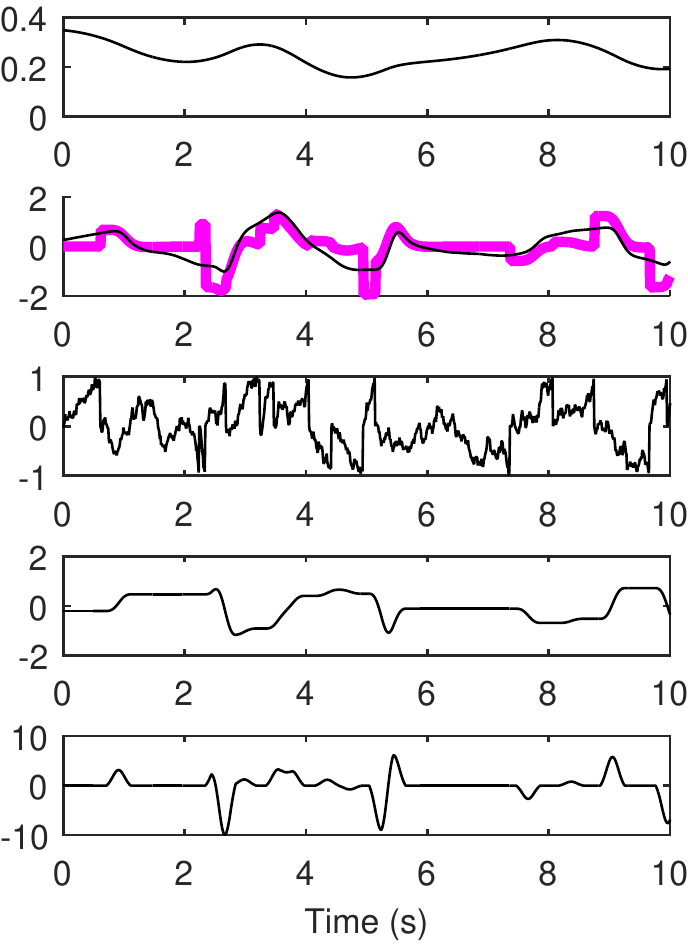}
	}
	\caption{Example simulations of the lane-keeping steering model driving on a straight road, with model parameters as in \tabref{ModelParameters}.}
	\label{fig:SimulationExamples}
\end{figure*}

To provide a further illustration of the proposed intermittent control framework, and a platform for testing its major assumptions, a model of ground vehicle steering will be described here. The full details will be developed over several sections below, but for illustration purposes some examples of the final model's time series behaviour are provided already in \figref{SimulationExamples}. Compared to the minimal example in \figref{MinimalExample}, note the effect, in panels (b) and (c), of introducing noise: Accumulator noise makes the adjustment timing less predictable, and motor noise causes a more inexact-looking steering profile, where $\Pp$ is generally \emph{not} equal to $\Prec$ just after the adjustment onset (cf.~\eqnref{PpWithMotorNoise}). The simulation in \figref{SimulationExamples}(c) also includes noise emulating random disturbances in the vehicle's contact with the road, in the form of a Gaussian disturbance to the vehicle's yaw rate, of standard deviation $\sigma\rmsub{R}$ and band limited to 0.5 Hz with a third-order Butterworth filter \citep{BoerEtAl2016}.

The steering model illustrated in \figref{SimulationExamples} uses the computational framework proposed here, with the perceptual control error quantity $P$ from \eqnref{SalvucciAndGray}, i.e.~the model is a generalisation to intermittent control of the steering model proposed by \cite{SalvucciAndGray2004}. The adopted control adjustment functions $\dot{G}$ and $G$ were the same as above; see \figref{GAndGdot}. This choice was based on the results by \cite{BenderiusAndMarkkula2014}, who showed that, across a wide range of real-traffic and driving simulator data sets, steering adjustments almost always followed a Gaussian-like rate profile, with average durations of about 0.4 s, encompassing about $\pm 2$ standard deviations of the Gaussian. As for $H$, note that again a sigmoidally decreasing function was used to generate the control error prediction $\Pp$. 

The plant model $S$ was on the general form of a linear so-called \inquotes{bicycle} model of lateral vehicle dynamics \citep{Jazar2008}:
\begin{equation}
\label{eq:LinearVehicleModel}
\left[
\begin{array}{c}
	\dot{v}\rmsub{y}(t) \\
	\dot{\omega}(t)
\end{array}
\right] 
=
\boldsymbol{A}_{2 \times 2} \left[ 
\begin{array}{c}
	v\rmsub{y}(t) \\
	\omega(t)
\end{array}
 \right] + \boldsymbol{b}_{2 \times 1}\delta(t), 
\end{equation} 
where $v\rmsub{y}$ is lateral speed in the vehicle's reference frame, $\omega$ is the rate of yaw rotation of the vehicle in a global reference frame, and $\delta$ is the steering wheel angle, i.e.~$C = \delta$. Here, the $\mathbf{A}$ and $\mathbf{b}$ matrices were obtained by fitting to observed vehicle response in two experiments with human drivers. 

These data sets of human steering, and how they have been analysed to (i) test framework assumptions and (ii) parameterise the model simulations shown in \figref{SimulationExamples}, will be described in \secref{TestingModelPredictions}. There, it will be shown that human vehicle steering in the studied data sets could be well described as a sequence of stepwise control adjustments, that the amplitude of individual adjustments could be explained using perceptual control error quantities at the time of adjustment onset, that the adjustment amplitudes could be even better explained if assuming that the humans responded to control error prediction errors rather than control errors per se, and finally that an accumulator model explained distributions of adjustment amplitude and timing better than did a threshold-based model. First, however, \secref{ReconstructionMethod} will introduce an analysis method that will be needed in the following.

\section{A simple method for interpreting sustained control as intermittent}
\label{sec:ReconstructionMethod}

Methods exist for decomposing shorter movement observations into a sequence of stepwise primitives, for example based on optimisation \citep{RohrerAndHogan2003, PolyakovEtAl2009} or high-order derivatives of the position signal \citep{FishbachEtAl2005}. Here, given our sustained control data with thousands of control adjustments, we adopt a considerably simpler method which is less exact, but also less computationally expensive and requiring only first-order derivatives.

For a given digitally recorded control signal with $N$ samples $C(j)$ taken at times $t(j)$, if one can estimate the times $t_i$ of control adjustment onset, one can use a discretized version of Eq.~(\ref{eq:ControlSummation}),
\begin{equation}
\label{eq:DiscreteSteeringSummation}
C(j) = C_0 + \sum_{i=1}^n \gtildei G(t(j) - t_i),
\end{equation}
to approximately reconstruct $C(j)$ as $n$ stepwise control adjustments with amplitudes $\gtildei$. By rewriting Eq.~(\ref{eq:DiscreteSteeringSummation}) as the overdetermined matrix equation
\begin{equation}
\boldsymbol{C} = \boldsymbol{G} \boldsymbol{g},
\end{equation}
where
\begin{equation}
\boldsymbol{C} = [C(1) \; \ldots \; C(N)]^T,
\end{equation}
\begin{equation}
\boldsymbol{G} = 
\left[ 
	\begin{array}{cccc}
		1 & G\left(t(1) - t_1\right) & \cdots & G\left(t(1) - t_n\right) \\
		\vdots & \vdots & & \vdots \\
		1 & G\left(t(N) - t_1\right) & \cdots & G\left(t(N) - t_n\right)
	\end{array}
\right]
\end{equation}
(i.e., a matrix with $N$ rows and $n + 1$ columns), and
\begin{equation}
\boldsymbol{g} = [C_0 \; \tilde{g}_1 \; \ldots \; \tilde{g}_n]^T,
\end{equation}
one can obtain a standard least-squares approximation of $\boldsymbol{g}$ using:
\begin{equation}
\hat{\boldsymbol{g}} = (\boldsymbol{G}^T\boldsymbol{G})^{-1} \boldsymbol{G}^T \boldsymbol{C},
\end{equation} 
or more efficient numerical techniques.

In order to estimate the times $t_i$ of adjustment onset, one can make use of the fact that if a signal is composed of intermittent discrete adjustments with sufficient spacing between them, then each adjustment will show up as an upward or downward peak in the rate of change of the signal (cf.~\figsref{MinimalExample} and \ref{fig:SimulationExamples}). Therefore, a simple approach to estimating the $t_i$  is to look for peaks in the control rate signal, after some appropriate amount of noise filtering, and define the steering adjustment onsets as occurring a time $T\rmsub{peak}$ before these peaks, where $\dot{G}($T\rmsub{peak}$)$ is the control rate maximum; i.e.~here $T\rmsub{peak} = \taum + \Delta T / 2$.

\section{Testing framework assumptions using human steering data}
\label{sec:TestingModelPredictions}

\subsection{Data sets}

To test the framework assumptions introduced here, two data sets of passenger car driving in a high-fidelity driving simulator were used:

(1) One set of 15 drivers recruited from the general public, performing routine lane-keeping on a simulated rural road, in an experiment previously reported on as Experiment 1 in \citep{KountouriotisAndMerat2016}. Here, only a subset of these data were used, by extracting the conditions with a straight road, no secondary task distraction and no lead vehicle. In total there were four segments of such driving per participant, each 30 s long. The average observed speed was 97 km/h.

(2) One set of eight professional test drivers performing a near-limit, low-friction handling task on a circular track (50 m inner radius) on packed snow. The task was to keep a constant turning radius, at the maximum speed at which the driver could maintain stable control of the vehicle. Each driver performed the task four times, and here 15 s were extracted from each such repetition, beginning at the start of the second circular lap, at which point drivers had generally reached a fairly constant speed (observed average 43 km/h). The motivation for including these data here was to study a more extreme form of lane-keeping, where driver steering is arguably operating in an \emph{optimizing} rather than a \emph{satisficing} mode \citep{Summala2007}. Three recordings where the driver lost control (identified as heading angle relative to circle tangent $> 10\degree{}$) were excluded.

In both experiments, the University of Leeds Driving Simulator was used. In this simulator, drivers sit in a Jaguar S-type vehicle cockpit with original controls, inside a spherical dome onto which visual input of 310\degree{} coverage (250\degree{} forward, 60\degree{} backward via rear view mirror) is projected. Motion feedback is provided by an eight degree of freedom motion system; a hexapod mounted on a lateral-longitudinal pair of 10 m rails. In both experiments, the steering wheel angle was recorded at a 60 Hz sample rate, with 0.1\degree{} resolution.

\subsection{Interpreting steering as intermittent control}
\label{sec:InterpretingSteeringAsIntermittent}

The computational framework developed in \secref{TaskGeneralModel} describes control as a sequence of stereotyped stepwise adjustments, with zero control change in between. Is it possible to understand the human steering in our data sets in this way? As a simple first indication, the fraction of time steps with zero change in steering wheel angle was indeed found to be rather large for both tasks: 45.8 \% for the circle task, and 91.1 \% for the lane-keeping; cf.~the plateaus in \figref{TustinExample}. 

\begin{figure}
	\subfloat[Lane-keeping]{
		\includegraphics[width=.48\columnwidth]{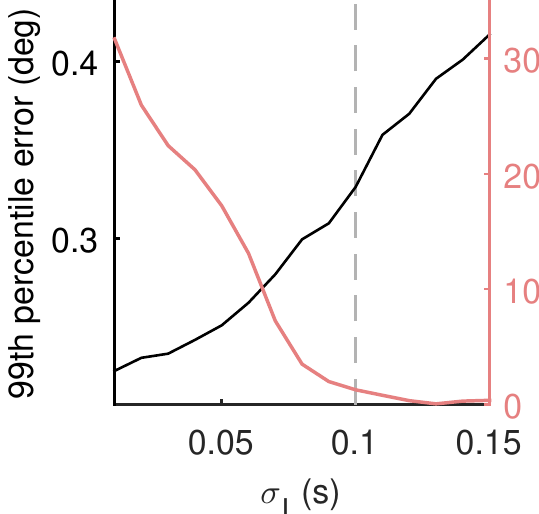}
	} 
	\subfloat[Circle task]{
		\includegraphics[width=.48\columnwidth]{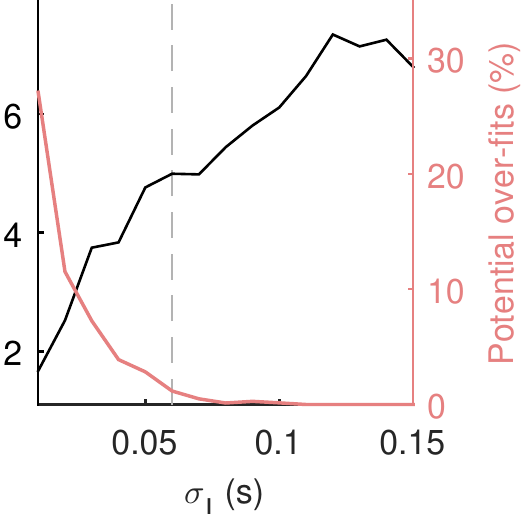}
	}
	\caption{The effect of low-pass filtering on the reconstruction of human steering as intermittent stepwise adjustments. Less filtering (lower $\sigmaI$) produces more exact reconstructions, but with a larger fraction of potentially over-fitted steering adjustments (see the text for details).}
	\label{fig:IntermittencyParameterVariation}
\end{figure}

To get a more complete answer, the method introduced in \secref{ReconstructionMethod} was applied, using the bell-shaped control adjustment $G$ described in \secref{ApplicationToSteering}. The noise filtering of the steering wheel signal, here achieved using a Gaussian-kernel averaging filter, does affect the outcome of this method, since a more heavily filtered signal will present fewer control rate peaks. Therefore, as illustrated in \figref{IntermittencyParameterVariation}, lower values of the filter kernel standard deviation $\sigmaI$ produced fits with larger numbers of steering adjustments and lower reconstruction error, here quantified in terms of 99th percentile of the absolute difference between recorded and reconstructed steering wheel angle. 

However, reconstructing with frequent adjustments also means that more of these are partially overlapping. It was found that this could produce unwanted effects, such as one fitted adjustment of large positive amplitude being followed by one large negative adjustment, producing a near-zero reconstructed steering angle. Such over-fitting tendencies were identified by comparing the peak steering wheel rate of the individual fitted adjustments to the observed steering wheel rate at the same points in time. These need not be identical, but when the fitted peak amplitude was more than 1.25 times larger than the observed steering rate peak, the adjustment was deemed a possible over-fit. The fraction of such adjustments are graphed against the right y axis in \figref{IntermittencyParameterVariation}. Based on these results, $\sigmaI$ was fixed at 0.1 s and 0.06 s for the lane-keeping and circle tasks, respectively.

\begin{figure*}
	\subfloat[Typical lane-keeping example.]{
		\includegraphics[width=.335\textwidth]{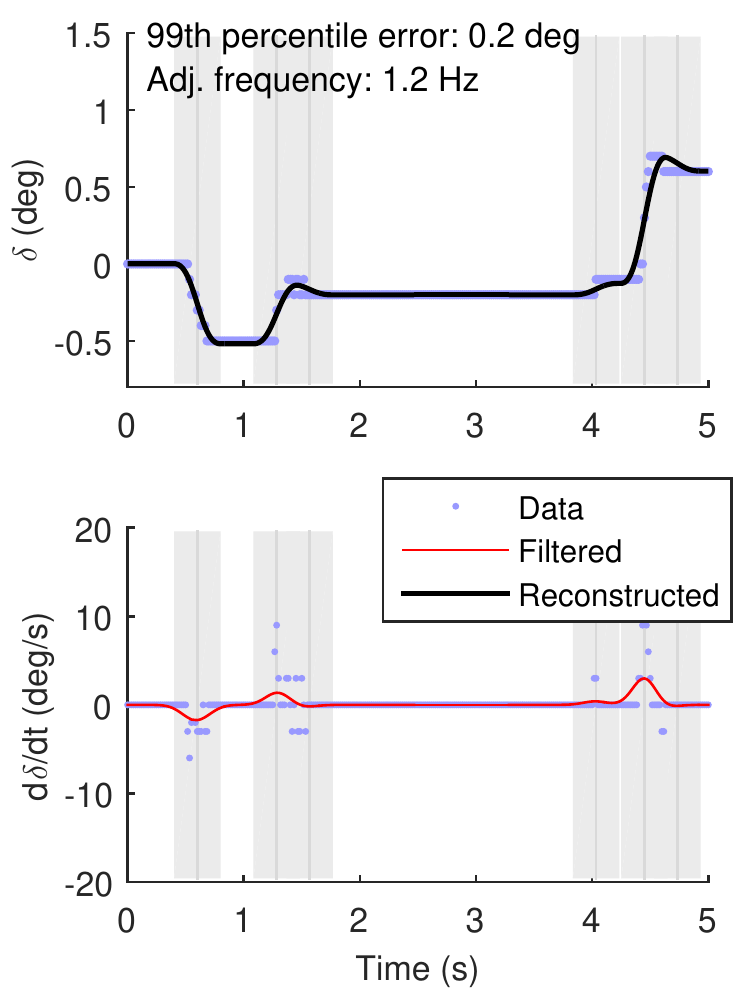}
	} 
	\subfloat[More frequent adjustments.]{
		\includegraphics[width=.315\textwidth]{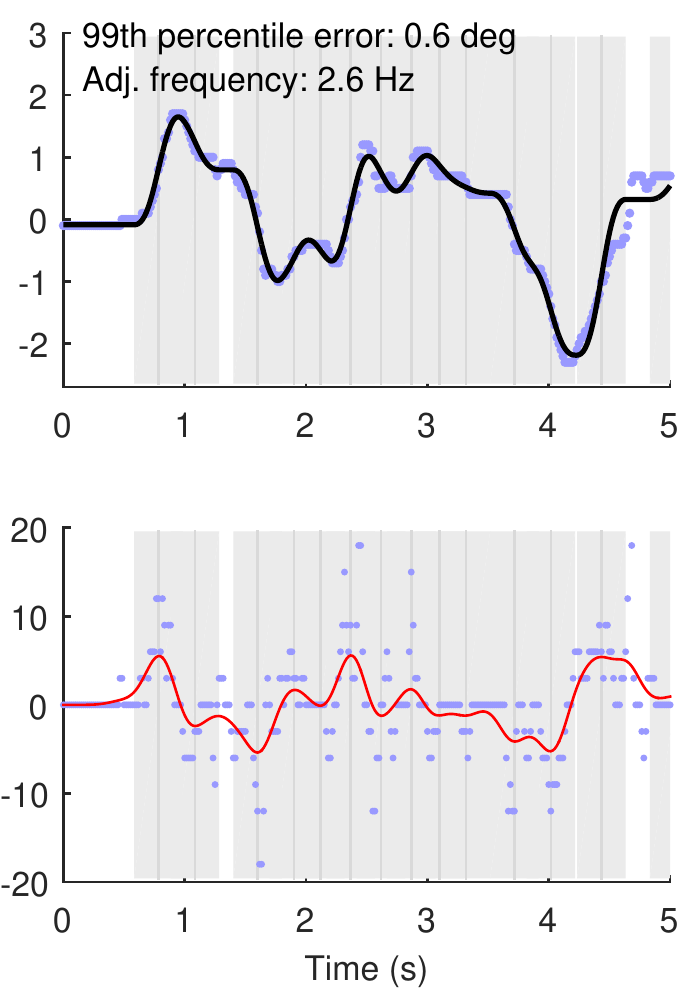}
	}
	\subfloat[Typical circle task example.]{
		\includegraphics[width=.325\textwidth]{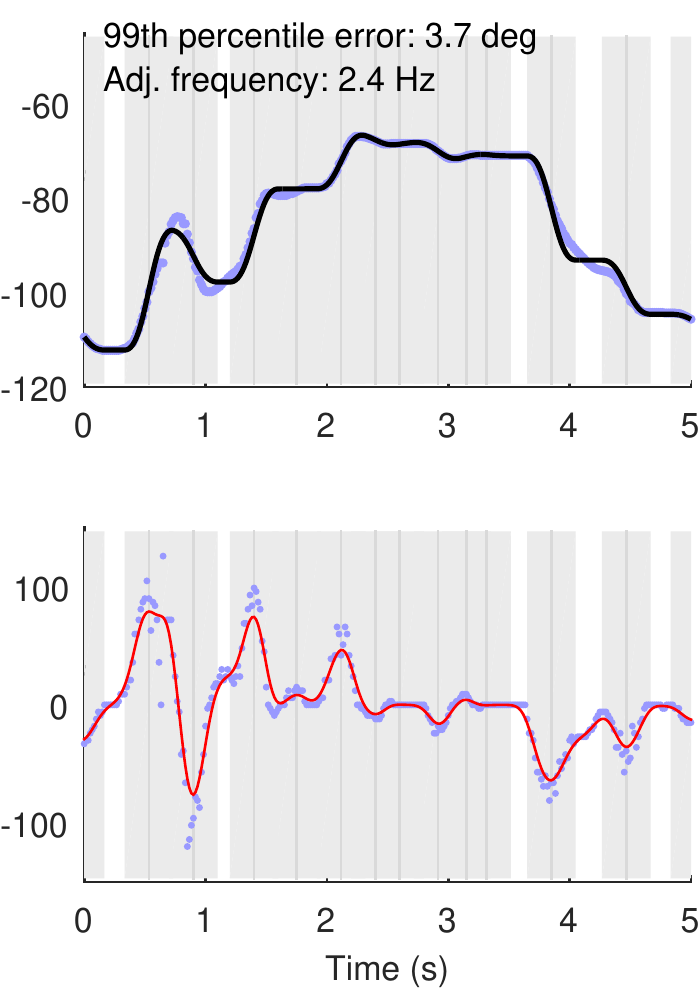}
	}
	\caption{Example reconstructions of the observed human steering as intermittent, stepwise control.}
	\label{fig:ExampleIntermittentReconstructions}
\end{figure*}

With these values for $\sigmaI$, the estimated adjustment frequencies, across the entire data sets, were 1.1 Hz and 2.0 Hz for the two tasks, a 98.2 \% and 96.6 \% compression compared to the original 60 Hz signals. As can be seen in \figref{IntermittencyParameterVariation}, 99th percentile reconstruction errors were 0.33\degree{} and 5.0\degree{} in the two tasks. These values were seemingly inflated somewhat by certain recordings with atypically large reconstruction errors. At the level of individual recordings, median reconstruction errors were 0.23\degree{} and 3.0\degree{}. \figref{ExampleIntermittentReconstructions} shows examples of reconstructions that are typical in terms of estimated adjustment frequencies and reconstruction errors, as well as one example lane-keeping recording with a higher estimated frequency of control adjustment, and a larger reconstruction error. 

Overall, these rather exact reconstructions using a small number of adjustments can be taken to suggest that something like intermittent stepwise control was indeed what drivers were making use of in these steering tasks. Such an interpretation seems qualitatively reasonable also from simply looking at the lane-keeping steering data, which, as mentioned, for the most part looked like \figref{ExampleIntermittentReconstructions}(a). Also the circle task steering, such as exemplified in \figref{ExampleIntermittentReconstructions}(c), had a decidedly staircase-like aspect. With examples such as the one shown in \figref{ExampleIntermittentReconstructions}(b), it is less qualitatively clear from the recorded steering signal itself that intermittent control might have been the case, but if one studies this plot closer (e.g.~supported by the vertical stripes in the figure), one can see why a reconstruction as a limited number of stepwise changes works also here: Basically, the control signal tends to always be either roughly constant (at 0 s, 1.3 s, 3.5 s, 4.8 s) or is changing upward or downward in a manner which can be understood as a single adjustment of about 0.4 s duration or shorter. Crucially, if control changes in the same direction for more than 0.4 s, it tends to do so with several identifiable peaks of steering rate (at 3-3.5 s and 3.6-4.2 s). A main cause of less exact reconstruction seems to be cases where two such peaks come close enough together to merge into one peak in the low-pass filtering (around 4.5 s in \figref{ExampleIntermittentReconstructions}(b) and around 0.6 s, 1.2 s, and 4 s in \figref{ExampleIntermittentReconstructions}(c)).

\subsection{Amplitude of individual steering adjustments}
\label{sec:AmplitudeResults}

The proposed framework also suggests that it should be possible to predict the control adjustment amplitudes from the control situation at adjustment onset, and more so than what it is possible to predict continuous rates of control change from the continuously developing control situation. To see whether this is the case here, we first consider a simplified, prediction-free version of \eqnref{IdealAdjustmentMagnitude}, where the expected value of control adjustment amplitude is
\begin{align}
\label{eq:POnlyAmplitudeModel}
g_i & = K' \Prec(t_i) = \{ K' = 1 \} = \Prec(t_i) \nonumber \\ 
 & = \knI \thetan(t_i - \taup) + \knP \thetandot(t_i - \taup) + \kf \thetafdot(t_i - \taup).
\end{align}
In \figref{Flowcharts}(b), this corresponds to the lower part of the model (\inquotes{Superposition of motor primitives}) being fed $\Prec$ directly instead of $\epsilon$. 
Note the similarity with the original, continuous \cite{SalvucciAndGray2004} control law in \eqnref{SalvucciAndGray}, which, with $K = 1$ and with the continuous model delay included, is:
\begin{equation}
\label{eq:SEtGInclDelay}
\dot{\delta}(t) = \knI \thetan(t - \taud) + \knP \thetandot(t - \taud) + \kf \thetafdot(t - \taud).
\end{equation}
This corresponds directly to the model in \figref{Flowcharts}(a).

Here, both the intermittent model in \eqnref{POnlyAmplitudeModel} and the continuous model in \eqnref{SEtGInclDelay} were fitted to the observed $g_i$ and $\dot{\delta}$, respectively, by means of a grid search, per driver, across all combinations of $\knI \in \{ 0, 0.01, ..., 0.20 \}$, $\knP \in \{ 0, 0.1, ..., 2 \}$, and $\kf \in \{ 0, 0.4, ..., 12 \}$ for the continuous model, and the same search ranges for \eqnref{POnlyAmplitudeModel}, but scaled by $\Delta T = 0.4$ s (cf.~\eqnref{KAndKPrimeRelationship}). The delay in \eqnref{SEtGInclDelay} was fixed at $\taud = 0.2$ s, after initial exploration suggested that values close to this one worked well across all drivers. For the intermittent model, the $g_i$ should correlate with the externally observed $P$ at a point $\taup + \taum + \Delta T/2$ before the peak of the observed adjustment; in this respect we here assumed $\taup + \taum + \Delta T/2 = 0.2$ s and did not vary these delays further. Also the preview times to near point and far point were fixed across drivers, again based on initial exploration, at 0.25 s and 2 s. 
 
\begin{figure*}
	\subfloat[Lane-keeping, CM]{
		\includegraphics[width=.325\textwidth]{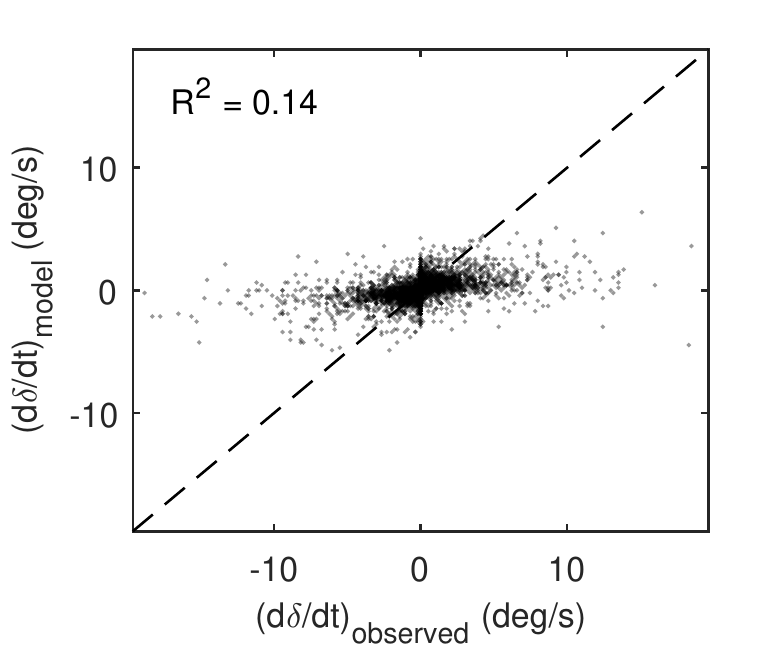}
	} 
	\subfloat[Lane-keeping, IM]{
		\includegraphics[width=.325\textwidth]{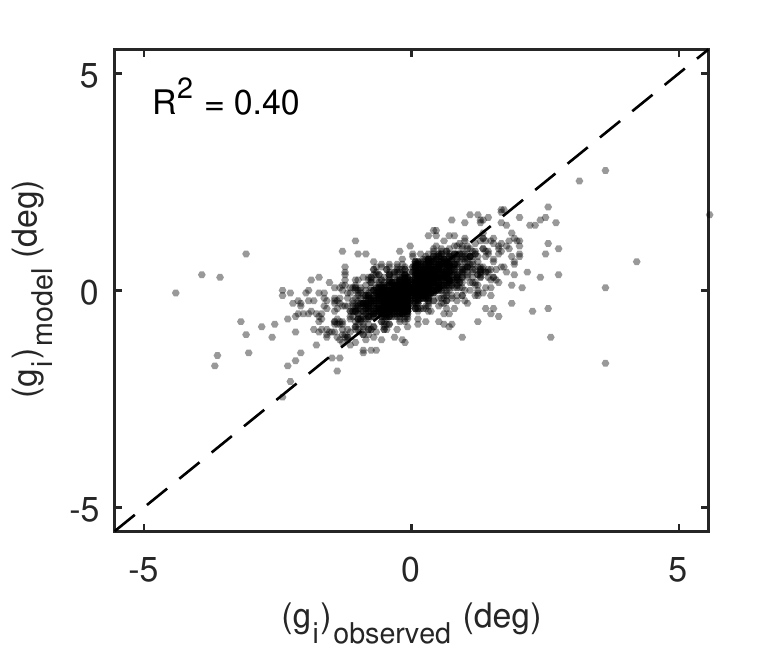}
	}
	\subfloat[Lane-keeping, PIM]{
		\includegraphics[width=.325\textwidth]{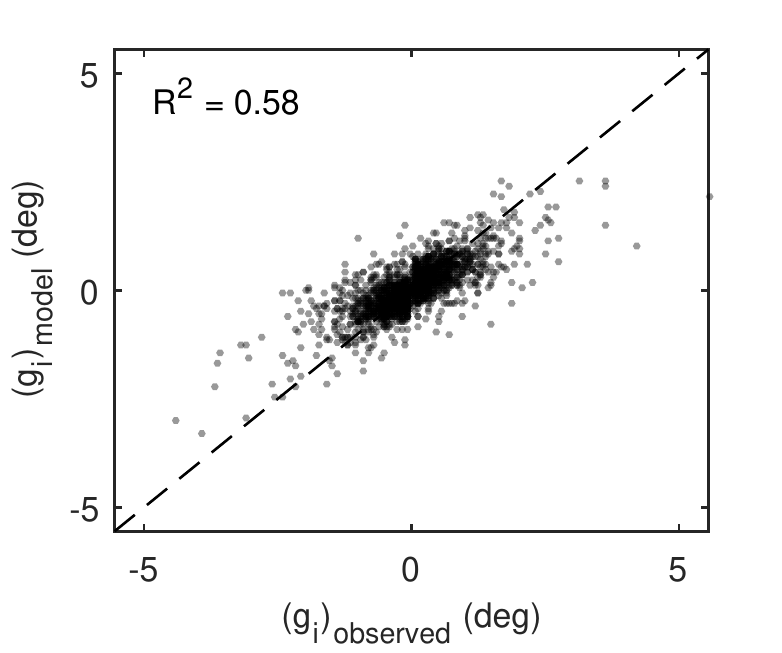}
	}
	\\
	\subfloat[Circle task, CM]{
		\includegraphics[width=.325\textwidth]{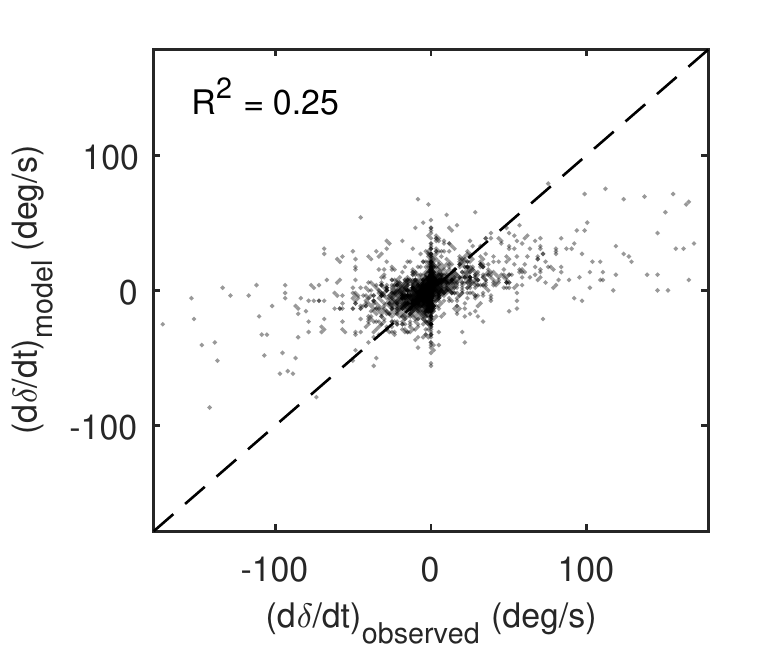}
	} 
	\subfloat[Circle task, IM]{
		\includegraphics[width=.325\textwidth]{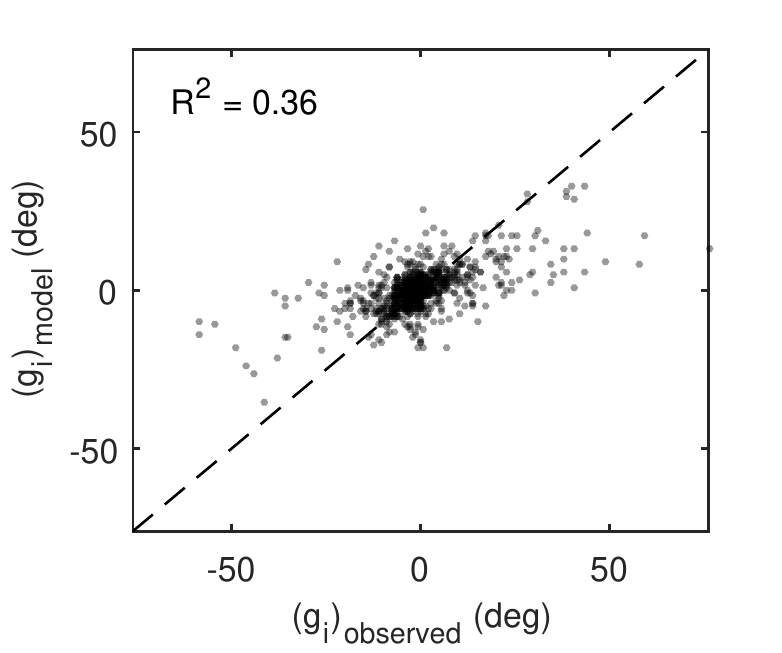}
	}
	\subfloat[Circle task, PIM]{
		\includegraphics[width=.325\textwidth]{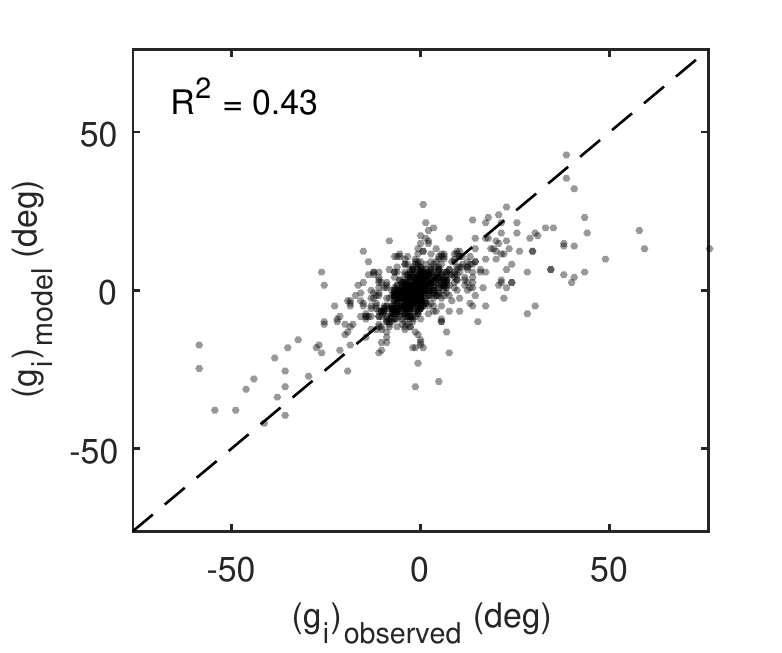}
	}
	\caption{Steering amplitude model fits; one continuous model responding to perceptual control error (CM; panels (a) and (d)), one intermittent model responding to perceptual control error (IM; panels (b) and (e)), and one intermittent model responding to errors in prediction of perceptual control error (PIM; panels (c) and (f)). The continuous and intermittent models predicted control rates ($\dot{\delta}$) and control adjustment amplitudes ($g_i$), respectively.}
	\label{fig:AmplitudeFittingResults}
\end{figure*}

\figref{AmplitudeFittingResults} shows, for both driving tasks, the entirety of observed and model-predicted control for the best-fitting gain parameterisations, for both the continuous model (panels (a) and (d)) and the intermittent model (panels (b) and (e)). 

\begin{figure*}
	\subfloat[]{
		\includegraphics[width=.335\textwidth]{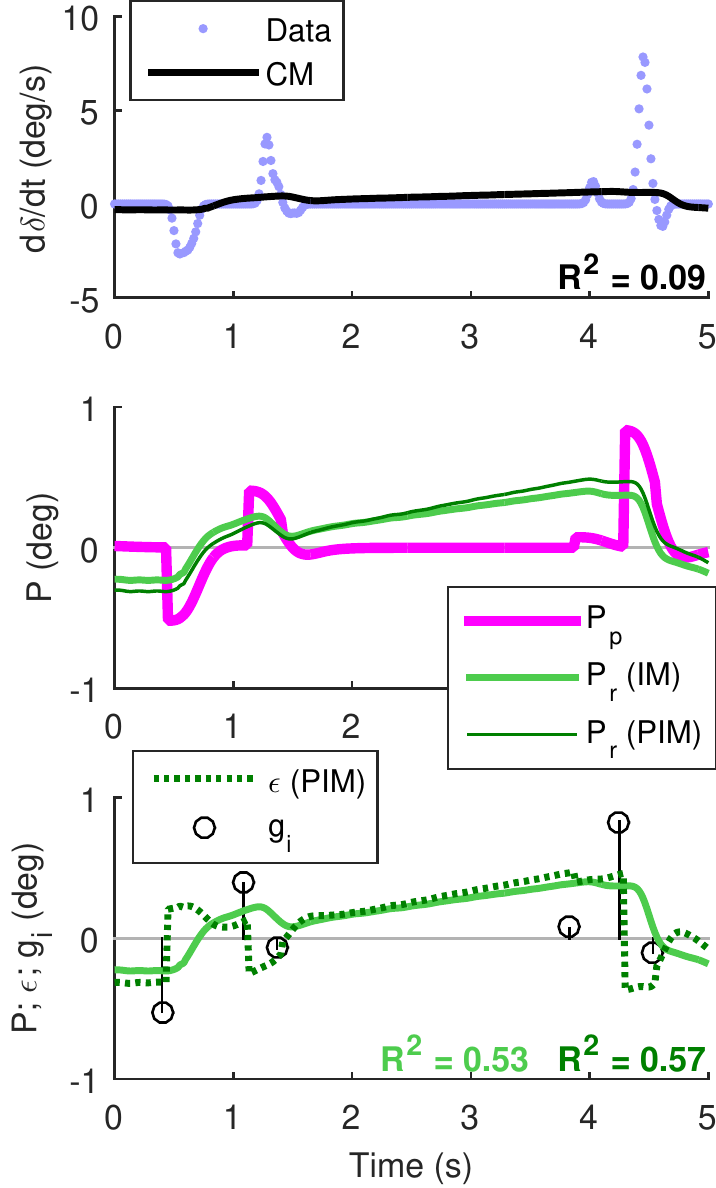}.pdf
	} 
	\subfloat[]{
		\includegraphics[width=.315\textwidth]{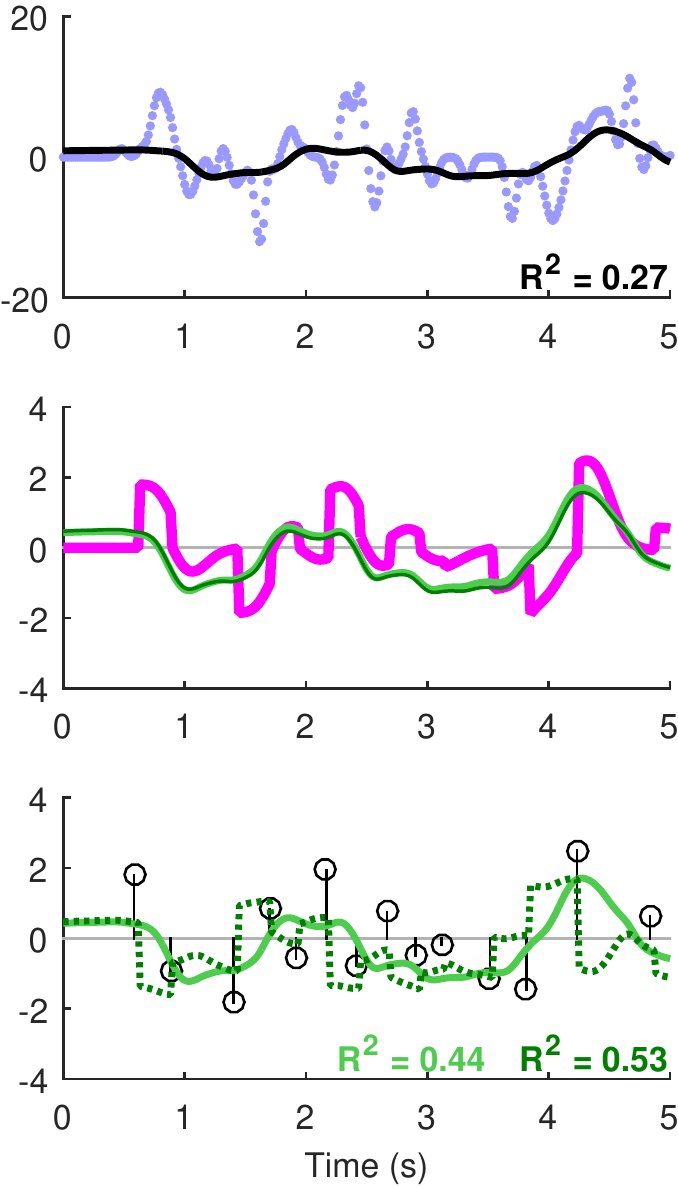}
	}
	\subfloat[]{
		\includegraphics[width=.325\textwidth]{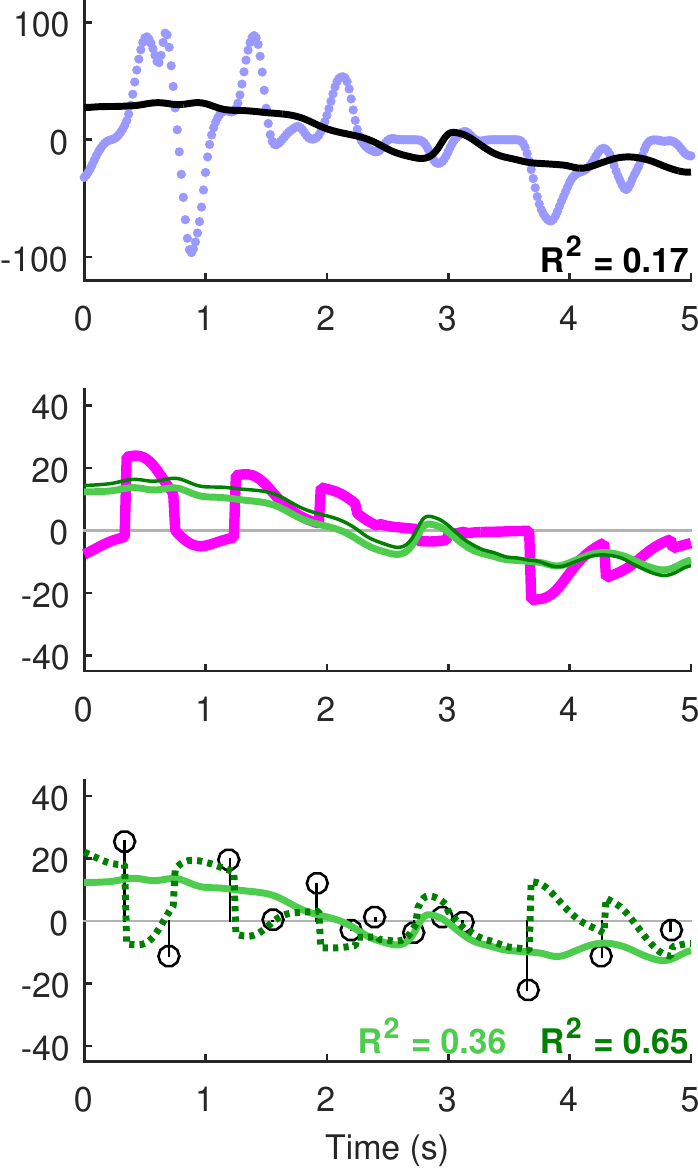}
	}
	\caption{Example illustrations of observed steering and fitted models of control amplitude, for the same three recordings as shown in \figref{ExampleIntermittentReconstructions}. The topmost panels show the continuous model (CM) fitted to observed steering rates. The middle plots show the $\Prec$ for both the intermittent and predictive inttermittent models (IM and PIM), and the $\Pp$ signal computed from the reconstructed adjustments. The bottom panels show the amplitude-predicting quantities of the two intermittent models, as fitted to the reconstructed adjustment amplitudes shown as vertical stems at the reconstructed times of adjustment onset (i.e.~not at the adjustment rate peaks); a perfect intermittent model would pass exactly through all circles.} 
	\label{fig:ExampleAmplitudeFits}
\end{figure*}

For the continuous model, note that the previously mentioned large fraction of time steps with zero change in the human steering is visible as vertical stripes in the middle of plots. The fitted gain parameters for this model are a compromise between not predicting too large steering rates for these stretches of zero control change, while nevertheless predicting non-zero steering rates of correct sign when the human actually is adjusting the steering; this is what is causing the data points in \figref{AmplitudeFittingResults}(a) and (d) to scatter at a flatter slope than the $y = x$ line that signifies perfect model fit. This compromise can be seen in more detail for the three example recordings in the top row of panels of \figref{ExampleAmplitudeFits}. As discussed in \secref{MinimalExample}, note that the continuous model behaviour looks like an average-filtered version of the steering rates, especially in panels (b) and (c) where there are many control adjustments.

For the intermittent model, the vertical stripes of data naturally disappear, as well as most of the flatness of the scatter. These results suggest that even though the model by \cite{SalvucciAndGray2004} was originally devised to explain continuous rates of steering change, it is actually better suited for explaining amplitudes of intermittent control adjustments; nicely aligning with the framework assumption being tested here. 

The bottom row of panels in \figref{ExampleAmplitudeFits} illustrates how the observed human control adjustment amplitudes $g_i$ relate to the variations over time of the parameter-fitted $\Prec$ quantity in \eqnref{POnlyAmplitudeModel}. As one would expect given the residual flatness of the scatter in \figref{AmplitudeFittingResults}, some of the above-mentioned model-fitting compromise remains; rather than hitting the observed $g_i$ directly, the fitted $\Prec$ tends to pass below (in absolute terms) the larger $g_i$, and above the smaller ones.

\subsection{Prediction of control errors}
\label{sec:PredictionResults}
Now, consider the full form of amplitude adjustment model proposed here, feeding $\epsilon$ rather than $\Prec$ to the motor control (the bottom part of \figref{Flowcharts}(b)):

\begin{align}
g_i & = K' \epsilon(t_i) = K'(\Prec(t_i) - \Pp(t_i)) \\
 & = \{ K' = 1 \} = \Prec(t_i) - \Pp(t_i)  \\
 & = \knI \thetan(t_i - \taup) +  \knP \thetandot(t_i - \taup) \ + \nonumber \\ 
 & \gatherindent \gatherindent \kf \thetafdot(t_i - \taup) - \Pp(t_i). \label{eq:AmplitudeModelWithPrediction}
\end{align}
If the framework proposed here is correct, \eqnref{AmplitudeModelWithPrediction} should explain adjustment amplitudes better than the prediction-free version in \eqnref{POnlyAmplitudeModel}.

\begin{figure}
  \includegraphics[width=\columnwidth]{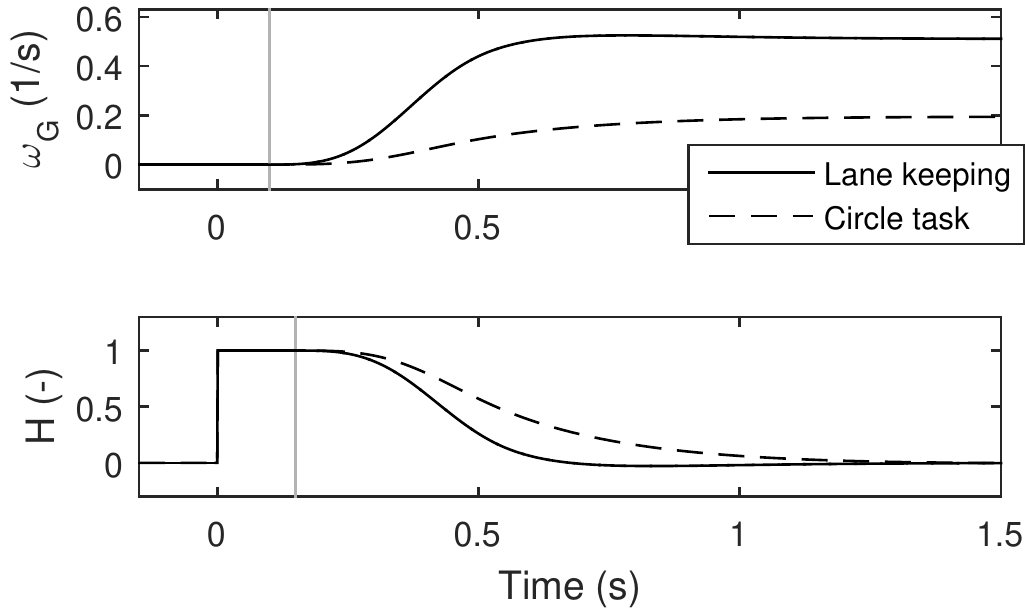}
	\caption{Top: Yaw rate responses to the sigmoidal steering adjustment profile $G$, of a linear vehicle model fitted to the two data sets of human steering. The vertical line indicates $\taum$. Bottom: Prediction functions $H$ for the two tasks, obtained using the yaw rate response profiles. The vertical line indicates $\taum + \taup$.}
	\label{fig:YawRateResponseAndH}
\end{figure}

To test whether this is the case, one needs to define suitable $H$ functions from which to build $\Pp$ (i.e., one needs to determine the \inquotes{superposition of prediction primitives} component in \figref{Flowcharts}(b)). Just as in \secref{MinimalExample}, besides the general requirements on $H$ set out in \eqnref{HRequirements}, we again make use of \eqnref{TheoreticalH}, suggesting that $H$ should describe how control errors decay when the controlled plant system responds to a control adjustment. In the case of ground vehicle steering, the plant $Y$ is the lateral dynamics of the vehicle, and as mentioned above these dynamics were here approximated using the linear model in \eqnref{LinearVehicleModel}. The $\boldsymbol{A}$ and $\boldsymbol{b}$ matrices of that equation were least-squares fitted to the two task data sets; \figref{YawRateResponseAndH} shows the yaw rate response $\omega\rmsub{G}(t)$ of the linear models thus obtained, when subjected to a steering input of the shape $G$ used here (as depicted in \figref{GAndGdot}).

Calculating exactly how an arbitrary $\Prec$ responds to a stepwise control adjustment $G$ is non-trivial, but it was found here that the following approximation of \eqnref{TheoreticalH} worked rather well in practice:
\begin{equation}
\label{eq:SteeringH}
H(t) = \begin{cases}
	0, & t \leq 0 \\
	1 - \omega\rmsub{G}(t - \taup) / S(v\rmsub{x}), & t > 0.
	\end{cases}
\end{equation}
where $S(v\rmsub{x})$ is the vehicle's steady state yaw rate response at longitudinal speed $v\rmsub{x}$, i.e.~for increasing $t$ $\omega\rmsub{G}(t) \rightarrow S(v\rmsub{x})$. This prediction function is shown in the bottom panel of \figref{YawRateResponseAndH}. In words, \eqnref{SteeringH} says that after applying a control adjustment $G$ to address a perceptual control error $\Prec$, this control error will over time fall towards zero with a profile that is the same as the profile of the vehicle's yaw rate response to $G$. This is only exactly correct if the actual control error is a pure yaw rate error (without heading or lane position errors). However, note in \figref{SimulationExamples}(a) that this $H$ nevertheless provides rather good prediction following most of the steering adjustments. For example, during the first rightward steering response to the leftward heading error, the prediction is exact while the adjustment is being carried out, and the far and near point rotations respond to the changing vehicle yaw rate. However, since the original error was not a yaw rate error, $P$ continues increasing above zero (which in turn prompts a sequence of stabilising steering adjustments to the right). \eqnref{SteeringH} thus serves as an example of what was speculated \secref{ControlErrorPrediction}; that also approximate predictions might in many control tasks be enough to allow successful control. Note that again $H$ takes the form of a sigmoid-like fall from one to zero.

Now, since we have fixed $K' = 1$, $\epstildei = \gtildei$, such that a $\Pp$ signal can be constructed using \eqnref{Pp} directly from the reconstructed $\gtildei$. Example prediction signals are shown in the middle row of panels in \figref{ExampleAmplitudeFits}. As shown in \figref{AmplitudeFittingResults}, using this $\Pp$ to fit the control gains in \eqnref{AmplitudeModelWithPrediction}, across the same parameter ranges as for \eqnref{POnlyAmplitudeModel}, yields improved fits, thus providing support for the framework assumption being tested here. It should be noted that these increases in model fit are not a result of introducing additional free parameters, but rather just of modifying the type of model function being used for fitting.

The bottom row of panels in \figref{ExampleAmplitudeFits} provides some further insight into the difference between models: When two adjustments follow each other with a short duration in between, the $\epsilon$ of the prediction-based model is often better than the prediction-free $\Prec$ at capturing the amplitude of the second adjustment, which tends to have a much smaller magnitude than $\Prec$, or even the opposite sign. The framework proposed here suggests that these small secondary adjustments occur because the preceding adjustments did not have quite the predicted effect. Especially for the lane-keeping task, this seemed to be happening more for some drivers than for others, and as one might expect it was to some degree related to frequency of steering adjustment. The three lane-keeping drivers for which the shift from \eqnref{POnlyAmplitudeModel} to \eqnref{AmplitudeModelWithPrediction} improved model fit the most, also had the three largest adjustment frequencies in the group.

\begin{figure}
	\includegraphics[width=\columnwidth]{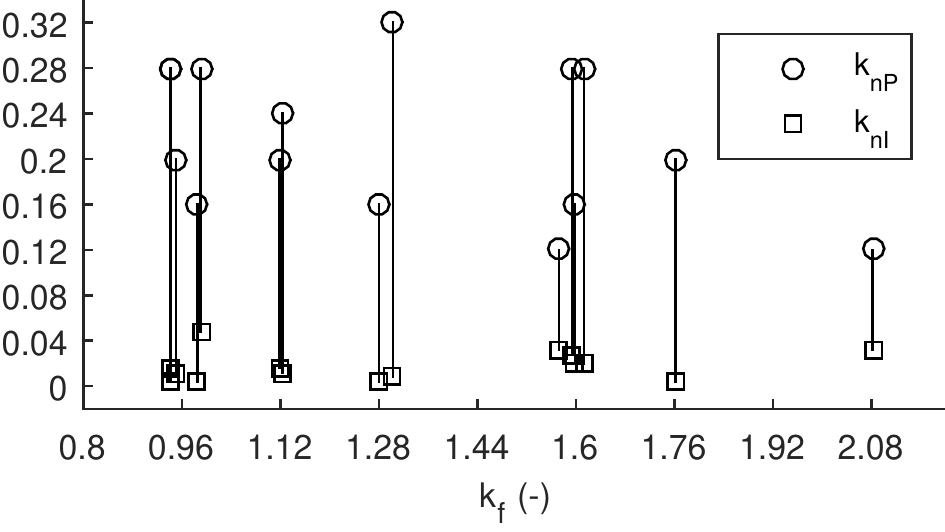}
	\caption{Best-fitting gain parameters for the prediction-extended \cite{SalvucciAndGray2004} model (\eqnref{AmplitudeModelWithPrediction}), when used to explain adjustment amplitudes in the lane-keeping data set. Each vertical line shows the fit for one driver. Slight random variation in $\kf$ has been added for legibility; the actual fitted values are the ones indicated on the $x$ axis.}
	\label{fig:SEtGGains}
\end{figure}

\figref{SEtGGains} shows the best-fitting gains obtained for the 15 drivers performing the lane-keeping task. Based on this figure, the gains $\knI = 0.02$, $\knP = 0.2$, and $\kf = 1.6$ were adopted for the example simulations in \figref{SimulationExamples}, and also for the further model fittings in the next section.

\subsection{Time between steering adjustments}
\label{sec:IAIVsAmplitudeResults}

A final theoretical prediction to be tested here is that the timing of observed adjustments should be better explained as generated by a process of evidence accumulation, such as set out in \eqnsref{GeneralisedAccumulator} or \eqref{eq:SimplifiedAccumulator}, than by control error thresholds or minimal refractory periods, such as adopted in existing frameworks and models of intermittent control \citep[e.g.~][]{MiallEtAl1993, GawthropEtAl2011, Benderius2014, JohnsAndCole2015, MartinezGarciaEtAl2016}. 

\begin{figure*}
\vspace{-.6cm}
\subfloat[Human drivers.]{
	\includegraphics[width=.32\textwidth]{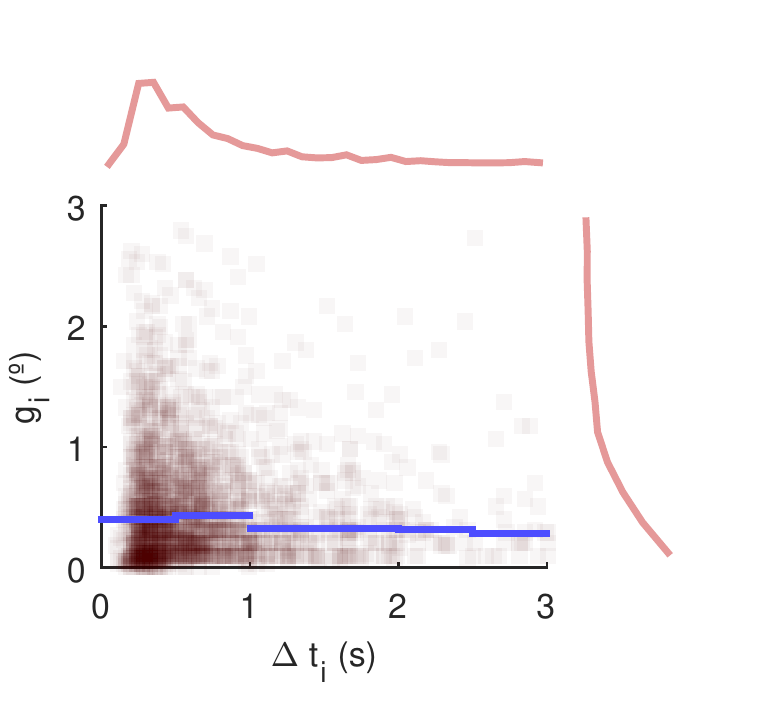}
	} 
\subfloat[Best-fitting threshold model.]{
	\includegraphics[width=.32\textwidth]{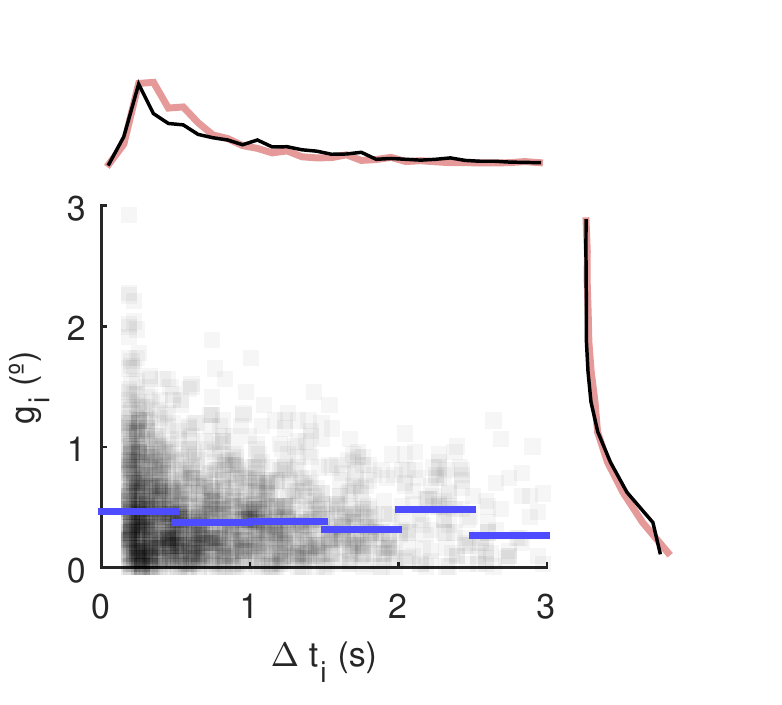}
	} 
\subfloat[Best-fitting accumulator model.]{
	\includegraphics[width=.32\textwidth]{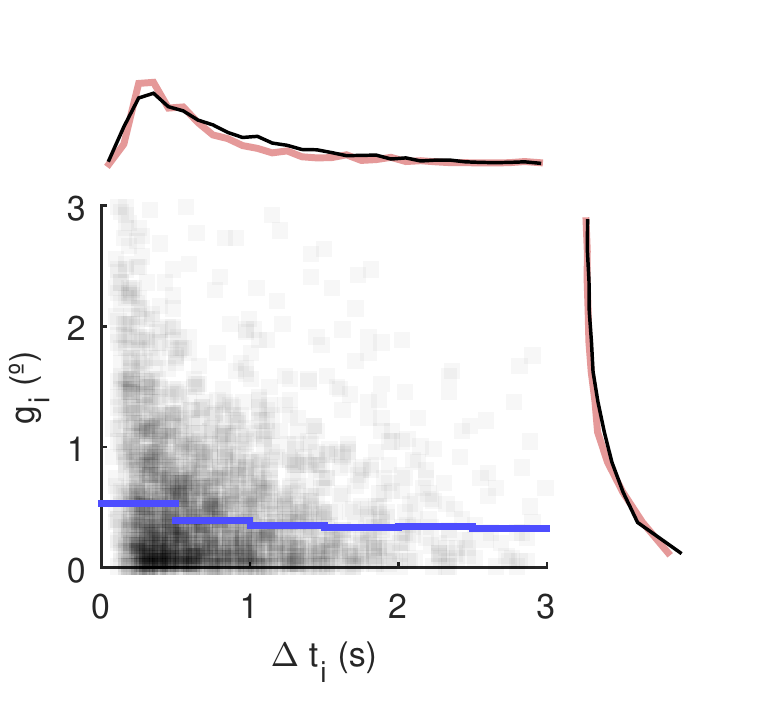}
	} 
\caption{Relationship between time $\Delta t_i$ since previous steering adjustment, and adjustment amplitude $\gtildei$, in the lane-keeping task. Each dot is one control adjustment, the curves show one-dimensional distributions, and the blue horizontal lines show median $\gtildei$ in bins of $\Delta t_i$. Panel (a) shows human steering data; panels (b) and (c) show computer simulations of best-fitting threshold-based and accumulator-based models, respectively.}
\label{fig:IAIVsAmplitude}
\end{figure*}

\begin{figure*}
\vspace{-.6cm}
\begin{center}
\subfloat[Human drivers, circle task.]{
	\includegraphics[width=.32\textwidth]{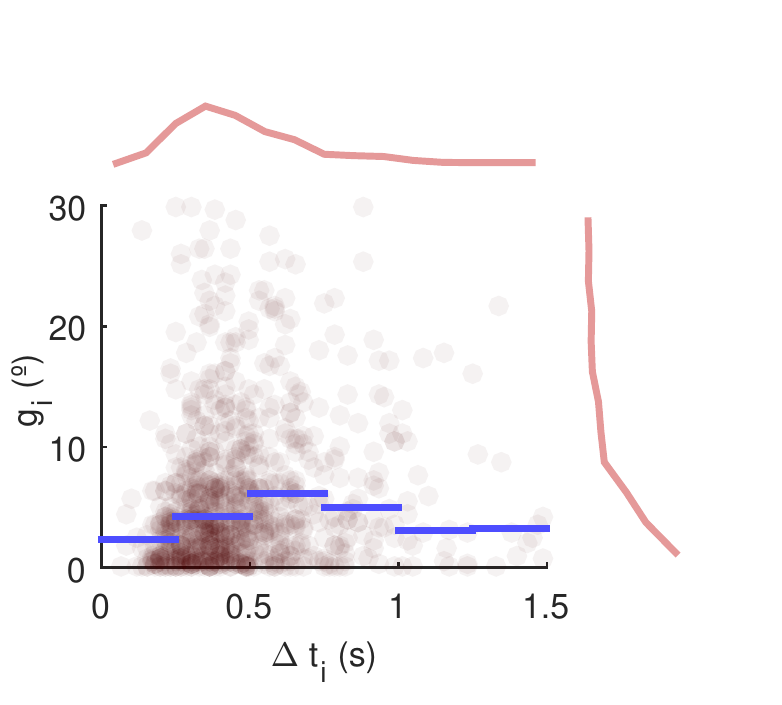}
	}
\hspace{.16\textwidth}
\subfloat[Lane-keeping model; road noise only.]{
	\includegraphics[width=.32\textwidth]{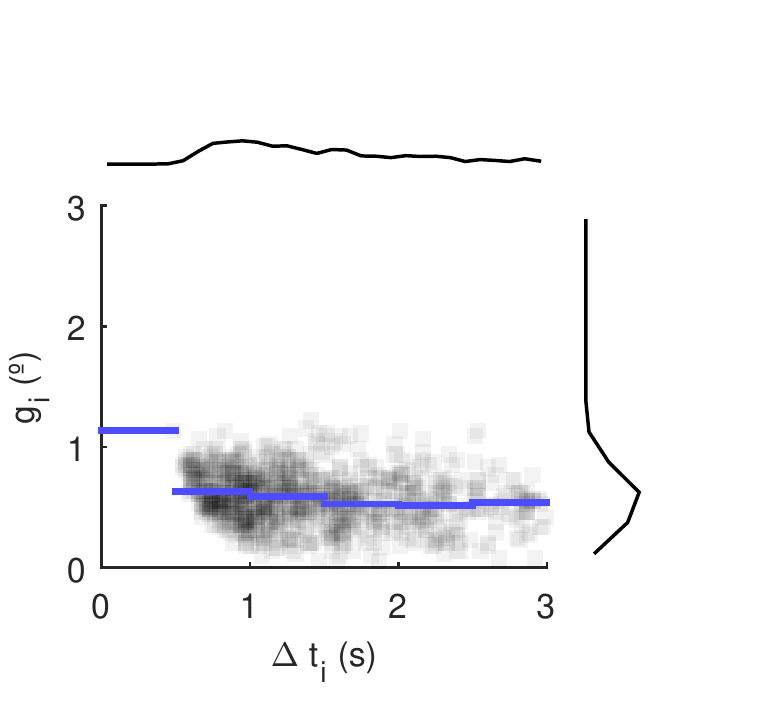}
	}
\\
\vspace{-.2cm}
\subfloat[\sigmaa = 0.4]{
	\includegraphics[width=.32\textwidth]{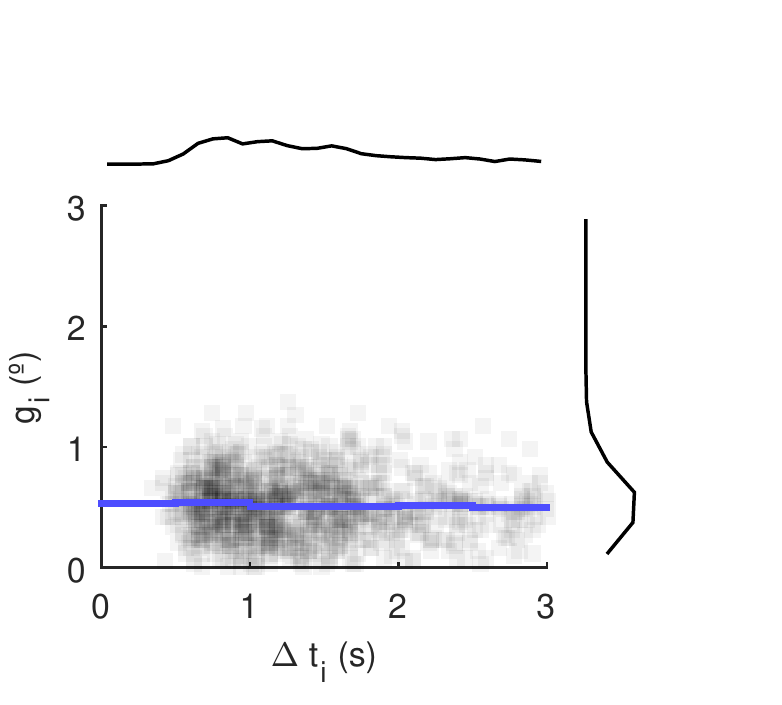}
	}
\subfloat[\sigmaa = 0.8]{
	\includegraphics[width=.32\textwidth]{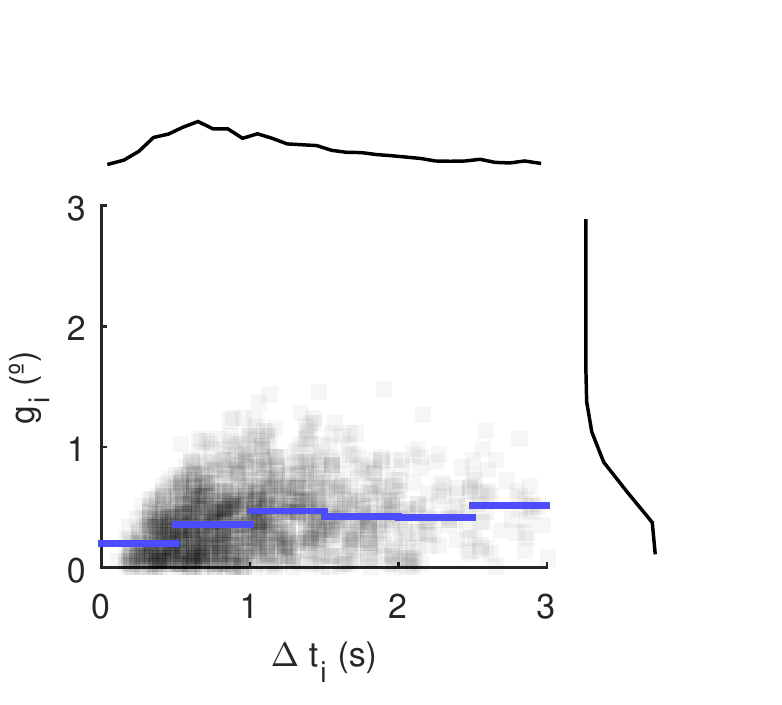}
	}
\subfloat[\sigmaa = 1.2]{
	\includegraphics[width=.32\textwidth]{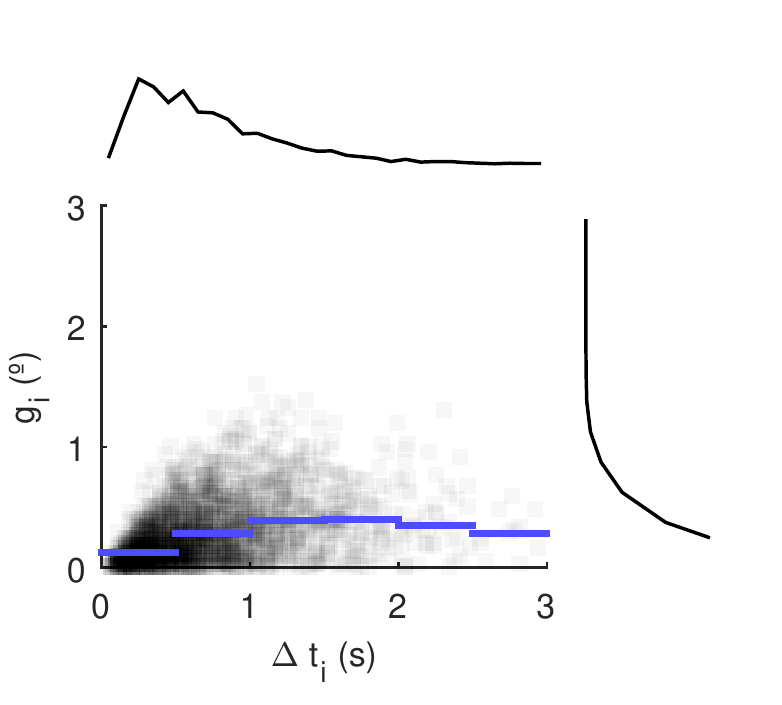}
	}
\\
\vspace{-.2cm}
\subfloat[\sigmam = 0.4]{
	\includegraphics[width=.32\textwidth]{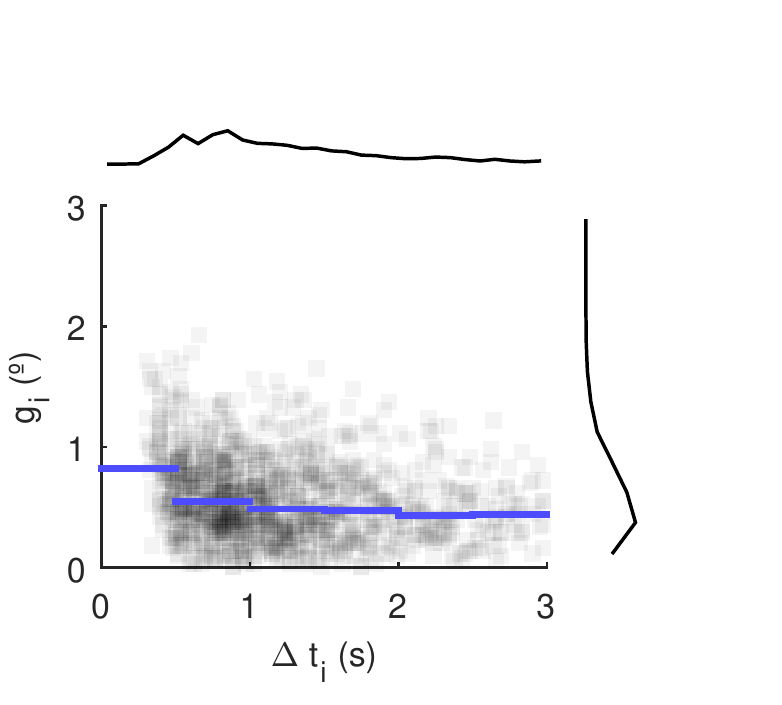}
	}
\subfloat[\sigmam = 0.8]{
	\includegraphics[width=.32\textwidth]{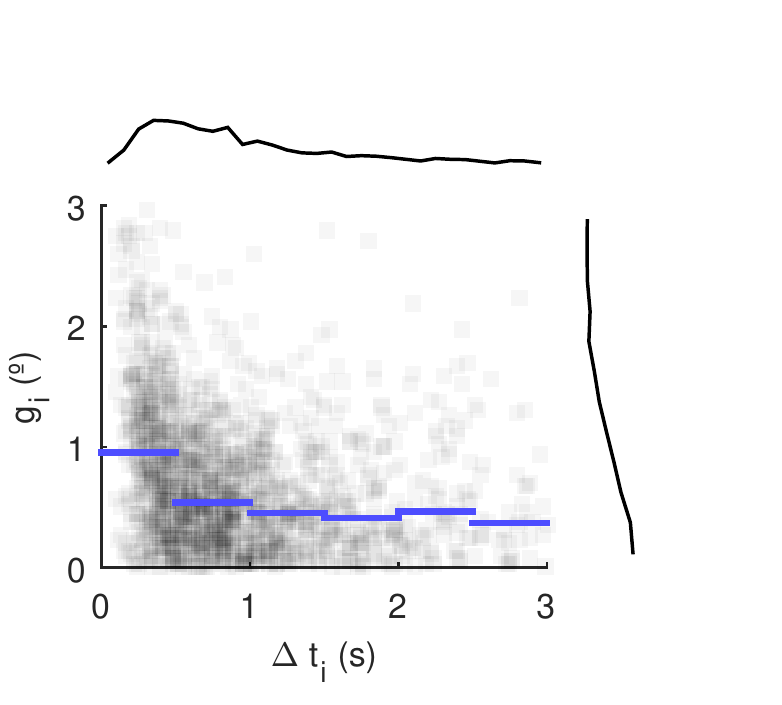}
	}
\subfloat[\sigmam = 1.2]{
	\includegraphics[width=.32\textwidth]{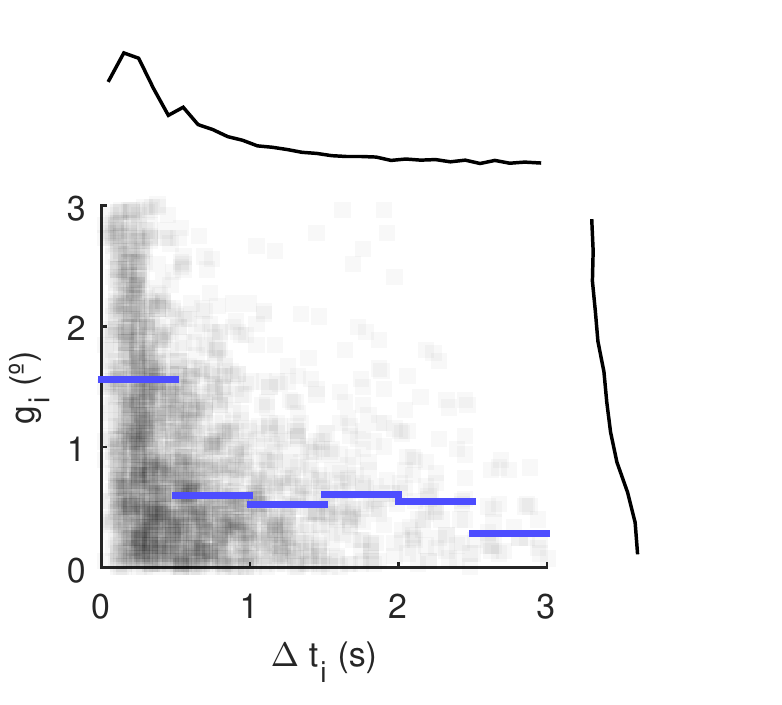}
	}
	\end{center}
	\caption{Further results on timing and amplitudes of steering; as in \figref{IAIVsAmplitude}. Panel (a) shows the human steering in the circle task, and panels (b-h) show the effects of varying noise levels in the best-fitting accumulator-based lane-keeping model shown in \figref{IAIVsAmplitude}(c). All simulations included road noise, at its fitted value $\sigmaR = 0.02$ rad/s. In panel (b) accumulator and motor noises ($\sigmaa$ and $\sigmam$) were set to zero, in panels (c-e) motor noise was zero and accumulator noise was varied around its fitted value (middle panel), and vice versa in panels (f-h).}
\label{fig:IAIVsAmplitudePartTwo}
\end{figure*}

\figsref{IAIVsAmplitude}(a) and \ref{fig:IAIVsAmplitudePartTwo}(a) show the distributions of not only adjustment amplitudes $\gtildei$ in the two data sets of human steering, across all drivers, but also the inter-adjustment interval $\Delta t_i \triangleq t_i - t_{i-1}$. In other words, these figures illustrate how the distribution of amplitudes varied with how much time had passed since the previous adjustment. Note that the distributions of $\Delta t_i$ (visible in collapsed form along the top of the panels) are roughly log-normal in character, skewed towards larger values, something which is typical of timings obtained from accumulator-based models \citep[e.g., ][]{BogaczEtAl2006}. 

Here, an approximate model-fitting of the lane-keeping data was carried out, using the \inquotes{typical} gain parameters obtained in \secref{PredictionResults} above, to see if fitting a single model to the data from all drivers would allow reproducing the general patterns seen in \figref{IAIVsAmplitude}(a). The remaining parameters of the steering model were grid searched, testing all combinations of the accumulator gain $k \in \{ 150, 200, ..., 400 \}$, the accumulator noise $\sigmaa \in \{ 0.4, 0.5, ..., 1.2 \}$, the motor noise $\sigmam \in \{ 0.2, 0.4, ... 1 \}$, and the road/vehicle noise $\sigmaR \in \{ 0.02, 0.03, 0.04, 0.05 \}$ rad/s. For each model evaluation, lane-keeping was simulated for the same amount of time as the human lane-keeping, i.e.~30 minutes of simulated driving. The model's steering adjustments were counted in bins with edges $\Delta t_i$ at $\{ 0, 0.2, 0.4, ... 6, \infty \}$ s, and for $\gtildei$ at \\ $\{ 0, 0.25, 0.5, ..., 3, \infty \}$ degrees, and the grid search identified the model parameterisation with minimum
\begin{equation}
\chi^2 = \sum_{j=1}^q \frac{(O_j - E_j)^2}{E_j + 1}
\end{equation}
where $O_j$ and $E_j$ are numbers of adjustment by humans and model in bin $j$, and $q$ is number of bins. This is standard chi-square minimisation distribution fitting, apart from the addition of one in the nominator, an approximate method to handle bins with $E_j = 0$. 

Also an alternative model was tested, intended to emulate typical assumptions of previous intermittent control models, as mentioned above. These previous models have been deterministic, and as such they are clearly unable to explain the data observed here. Therefore, an extended stochastic formulation was used: Instead of accumulating prediction error $\epsilon$, this model triggered new adjustments when time since previous adjustment exceeded $\Delta\rmsub{min}$ and $\epsilon + \nu\rmsub{t} \geq \epsilon_0$, where $\nu\rmsub{t}$ is Gaussian noise with zero mean and standard deviation $\sigma\rmsub{t}$, and $\epsilon_0$ a threshold parameter. This model was grid searched across all combinations of \\
$\sigma\rmsub{t} \in \{ 0.06, 0.12, ..., 0.52 \}$ degrees\footnote{From 0.001 to 0.009 radians.},
$\epsilon_0 \in \{ 0.2, 0.4, ..., 1.2 \}$ degrees, $\Deltamin \in \{ 0, 0.1, 0.2 \}$ s, and $\sigmam$ and $\sigmaR$ across the same ranges as for the accumulator model. 

For both models, the best solutions from the grid searches were optimised further using an interior point algorithm (Mathworks MATLAB function \emph{fmincon}). 

The best fits obtained are shown in \figref{IAIVsAmplitude}(b) and (c), with a lower $\chi^2 = 451$ (i.e., a better fit) for the accumulator model than for the threshold model, $\chi^2 = 593$, despite the threshold model having one more free parameter. The main shortcomings of the threshold model seemed to be (i) a tendency to produce a majority of control adjustments just after the $\Deltamin$ duration, thus not generating a very lognormal-looking distribution of $\Delta t_i$, and (ii) a failure to account for those observed data points which had simultaneously large $\Delta t_i$ and $\gtildei$. The fitted values for the accumulator model were used when generating the example simulations in \figref{SimulationExamples}. The full list of all parameter values used in those simulations are provided in \tabref{ModelParameters}.

\begin{table}
\renewcommand{\arraystretch}{1.3}
\caption{Parameter values used for the lane-keeping steering model simulations in \figsref{SimulationExamples}, \ref{fig:IAIVsAmplitude}, and \ref{fig:IAIVsAmplitudePartTwo} (except where otherwise indicated in those figures). 
}
\label{tab:ModelParameters}
\centering
\begin{tabularx}{\columnwidth}{llX}
\hline
\bfseries Parameter & \bfseries Value & \bfseries Obtained from \\
\hline
$\taup$ & 0.05 s & Literature; see \secref{MinimalExample}. \\
$\taum$ & 0.1 s  & -''- \\
$\taun$ & 0.25 s & Exploratory fitting of steering amplitude models; see \secref{AmplitudeResults}.	\\
$\tauf$ & 2 s    & -''-	  \\
$\knI$  & 0.02   & Fitting of \eqnref{AmplitudeModelWithPrediction} to observed steering amplitudes; see \secref{PredictionResults} \\
$\knP$  & 0.2 s  & -''- \\
$\kf$   & 1.6 s  & -''- \\
$k$     & 200    & Fitting of accumulator model (\eqnref{SimplifiedAccumulator}) and noise magnitudes, to observed timing and amplitudes of steering; see \secref{IAIVsAmplitudeResults}. \\
$\sigmaa$ & 0.8 a.u. & -''- \\
$\sigmam$ & 0.8  & -''- \\
$\sigmaR$ & 0.02 rad/s & -''- \\
\hline
\end{tabularx}
\end{table}


Panels (b) through (h) of \figref{IAIVsAmplitudePartTwo} provide a closer look at how the accumulator-based model's behaviour varies when the different noise magnitudes are varied. In panel (b), note how, in the the absence of any accumulator or motor noise, adjustments become infrequent. This is because they are triggered solely by noise-free accumulation of control errors, which tend to be small due to the noise-free control (with gains fitted to the human steering) being rather well-attuned to the vehicle (cf.~\figref{SimulationExamples}(a)). 
A pattern of decreasing $\gtildei$ with increasing $\Delta t_i$, observable for human steering in both tasks with $\Delta t_i > 0.5$ s, is clear already in this simplified form of the model. This is a somewhat counterintuitive consequence of accumulation-based control \citep{Markkula2014}; integration of a small quantity over a long time is the same as integration of a large quantity over a short time\footnote{Also the noise-extended threshold model can give this qualitative behaviour, however in this case by the logic that with a small control error it takes a longer time, on average, before the noise happens to be large enough for threshold-passing.}. 

When adding and increasing accumulator noise (panels (c) through (e)), adjustments become more frequent, and smaller $\Delta t_i$ start occurring. At these lower $\Delta t_i$, there is now the opposite pattern of increasing $\gtildei$ with increasing $\Delta t_i$. This happens in the model because the earlier the noise happens to push the accumulator above threshold, the smaller the control error to respond to will be, on average. Interestingly, this sort of pattern can be seen clearly in the human steering in the circle task (panel (a)). If, instead of accumulator noise, we add and increase motor noise (panels (f) through (h)), we see that this is another way of producing small $\Delta t_i$, in this case because ill-attuned adjustments soon trigger additional, corrective adjustments. Here, since large motor mistakes will be detected more quickly, the smaller $\Delta t_i$ are here instead associated with larger $\gtildei$, thus counteracting the above-mentioned effect of accumulator noise.

\section{Discussion}
\label{sec:Discussion}

Below, some relevant existing accounts of sensorimotor control will first be enumerated and briefly contrasted with what has been proposed here. Then, a series of subsections will engage in more detail with some specific topics for discussion.

\subsection{Related models and frameworks}

As mentioned in the Introduction, Gawthrop and colleagues have also presented a task-general framework for intermittent control \citep{GawthropEtAl2011, GawthropEtAl2015}. What has been proposed here aligns well with their emphasis on possible underlying control intermittency even in cases where the overt behaviour is seemingly continuous in nature. However, at the level of actual model mechanisms, the two frameworks are rather different, with Gawthrop et al starting out from an optimal control engineering perspective whereas we have put more focus on adopting concepts from psychology and neurobiology: zero-order or system-matched holds versus motor primitives; explicit inverse and forward system models versus perceptual heuristics and corollary discharge-type prediction primitives; error deadzones and minimum refractory periods versus evidence accumulation. 

Another task-general framework has been derived from the \emph{free-energy principle}, which suggests that minimisation of free energy, or roughly equivalently minimisation of prediction error, is the fundamental governing principle of the brain
\citep{Friston2005, Friston2010}.
From this mathematical framework, Friston and colleagues have derived models of sensorimotor control as \emph{active inference} \citep{FristonEtAl2010, FristonEtAl2012ActiveInference, PerrinetEtAl2014}, but these have focused on continuous rather than intermittent control. The active inference framework, like ours, describes motor action as being generated to minimise sensory prediction errors, and sensorimotor control as near-optimal without being directly based on engineering optimal control mechanisms. However, these active inference models have not explicitly included notions of superpositioned ballistic motor primitives, or evidence accumulation to decide on triggering such primitives. In our understanding, such mechanisms should be obtainable as special cases of the more generally formulated active inference theory; our argument here is that these might be useful special cases to consider.

In contrast, as mentioned in \secref{ConceptualOverview}, some researchers focusing specifically on motor control have proposed superposition of sequences of motor primitives as a main feature of their conceptual frameworks \citep{HoganAndSternad2012, Karniel2013}, but so far without developing these into full computational accounts. Others have focused on how the primitives themselves might be constructed using underlying dynamical systems formulations \citep{IjspeertEtAl2003, SchaalEtAl2007}; a description one level below the one we have adopted here. There is also a related, vast literature on  neuronal-level models of how individual saccadic eye movements are generated \citep[e.g., ][]{GirardAndBerthoz2005, RahafroozEtAl2008, DayeEtAl2014}. Overall, these motor-level accounts suggest that the kinematic motor primitives considered in the present framework are not truly ballistic, in the sense that there is a closed control loop to support their successful motor completion. However, from a higher-level perspective it might still be correct to consider them ballistic, in the sense that once they are initiated, they are not further affected by how the perceptual situation which triggered them continues to evolve. 

There are also task-specific models of sensorimotor control sharing some of the present framework's assumptions. The task of reaching towards a target has for example been modelled as a superposition of two non-overlapping bell-shaped speed pulses by \cite{MeyerEtAl1988}, or as an arbitrary number of pulses with possible pairwise overlap by \cite{BurdetAndMilner1998}. Both of these models allow variable-duration primitives, and the latter model also includes provisions for uncertain estimation of predicted final amplitude of an ongoing primitive, in a manner that is related but not identical to the prediction error-based control used here. A more direct analogue exists in models of smooth pursuit of moving targets with the eyes, where the Smith Predictor type approach has long been used \citep[][and the same is actually true also for the above-mentioned models of individual saccades]{RobinsonEtAl1986, KettnerEtAl1997, GrossbergEtAl2012}, but these models are instead continuous in nature. Among the models of car steering as intermittent control, the ones by \cite{RoyEtAl2008} and \citep{JohnsAndCole2015} are more similar to the \cite{GawthropEtAl2011} framework than to ours, whereas the models by Gordon and colleagues \citep{GordonAndSrinivasan2014, GordonAndZhang2015, MartinezGarciaEtAl2016} do make use of steering adjustment primitives, but in a hybrid intermittent-continuous control scheme. The model by \cite{Benderius2014} uses motor primitives and perceptual heuristics, but not sensory prediction or evidence accumulation. The only other car steering model that hasn't used error deadzones is the one by \cite{BoerEtAl2016}, who used a just noticeable difference mechanism.

The overall impression is that the level of description we have adopted places our framework somewhere in the middle with  respect to these existing models. We are arguably one step closer to the neurobiology than the \cite{GawthropEtAl2011} framework and the existing car steering models, and one step further away from the neurobiology and from detailed behavioural-level knowledge than some of the models of manual reaching or eye movements. 
One topic of discussion in the sections to follow below will be how these higher-level and lower-level accounts might possibly benefit from adopting some of the ideas proposed here.

\subsection{Evidence accumulation in sensorimotor control}

To the best of our knowledge, no prior models have adopted the idea that evidence accumulation is involved in sustained sensorimotor control, to decide on when to change the current control by for example triggering a new open-loop control adjustment. This hypothesis seems a very natural one to explore given the large amount of empirical support for accumulation-type models in the context of single decision perceptual-motor tasks. What has been proposed here is that sustained sensorimotor control can be regarded as a sequence of such decisions. 

More specifically, we have proposed that the rate of accumulation towards the decision threshold might scale with control error prediction error. This provides an interesting possible answer to the long-standing open question whether control intermittency is caused by minimal refractory periods or to error deadzones, or both. For example, \cite{MiallEtAl1993} found that their data supported neither hypothesis completely, and \cite{vandeKampEtAl2013} reported evidence for a refractory period that varied with the order of the control task. In effect, accumulation of prediction error (or even of just control error, without predictions) will result in both (i) mandatory refractory pauses between control actions and (ii) control error magnitudes at which control actions will most typically be issued, but both of these will vary with the specifics of the control situation leading up to the adjustment, and quite naturally also with the task itself (as between the lane-keeping and circle steering tasks studied here). Furthermore, with noise included in the accumulation process, this type of model also provides a natural means of capturing the inherent stochasticity in control action timing. 

Given the above argument, it is interesting to note that our approximate model-fitting analyses tentatively favoured the accumulator-based account over the more conventional, threshold-based model (even when extended to a stochastic formulation). These findings should be complemented with more detailed analyses and targeted experiments. Such investigations could benefit from considering not only behavioural but also neuroimaging data, to possibly look for direct traces of any ongoing evidence accumulation \citep[see, e.g.,][]{WerkleBergnerEtAl2014}. One specific assumption in the present framework that would merit testing with both behavioural and neuroimaging approaches is the currently assumed resetting of the accumulation to zero immediately after each control adjustment.

%
%
%

\subsection{Different types of open-loop primitives}
\label{sec:PrimitiveTypesDiscussion}

The motor primitives we have considered here have been of a rather sinmple nature: stepwise changes of position, all of the same basic shape and duration regardless of amplitude. In car steering specifically, this approach aligns with a previous report of amplitude-independence in steering adjustments \citep{BenderiusAndMarkkula2014}, and it was also sufficient, here, for making the point that the car steering data could be much better understood as a sequence of such steps than as continuous control. However, if one wanted to apply the computational framework proposed here to other tasks (including car steering in a more general sense than lane-keeping or circle-tracking), one would most probably want to consider a wider variety of motor primitives.
 
Already at the level of simple stepwise position changes, it is clear that humans can adapt the duration of their limb movements to the requirements of the task at hand \citep{Plamondon1995}. Even within the same visuo-manual joystick tracking task ,\cite{HannetonEtAl1997} observed stepwise adjustment behaviour where smaller amplitude adjustments were performed faster. Visual inspection suggests that this latter phenomenon might actually be occurring also in the present car steering data sets (see, e.g., the small adjustment at 4 s in \figref{ExampleIntermittentReconstructions}(a)), but if so possibly at amplitudes which would require higher-resolution steering angle measurements to properly characterise.  

Also wider classes of kinematic motor primitives have been proposed. \cite{HoganAndSternad2012} suggested that in addition to stepwise kinematic submovements, a task-general account of motor control should also include primitives for kinematic \emph{oscillations} as well as \emph{impedances}. The car steering models by Gordon and colleagues propose that a higher-level set open-loop primitives is constructed from the simple stepwise adjustments; e.g.~one, two or three opposing steps in sequence to achieve a desired adjustment of either vehicle yaw rate, yaw angle, or lateral position, respectively \citep{GordonAndSrinivasan2014, MartinezGarciaEtAl2016}\footnote{They also propose an interesting method for automated identification of this dictionary of steering primitives, allowing a more powerful, but also more complex, signal reconstruction than the method we have used here.}. It has indeed been proposed that learning to construct finely task-attuned higher-level kinematic primitives in this type of manner might be an important role of the motor system \citep{Giszter2015}. A possible special case that would seem useful in many tasks, but that we have not seen mention of in the literature, would be a constant-rate primitive, e.g.~constructed from a rapid sequence of partially overlapping position changes. Interestingly, at least one early eye movement researcher described smooth pursuit eye movements as intermittent adjustments of movement rate \citep{Westheimer1954}.

On the motor control side of our framework, expanding to a larger number of kinematic primitives is straightforward; one would simply need to create a set of functions $G$ defining these primitives, and an associated set of prediction functions $H$. What would require some more thought is the decision-making mechanism, which would then no longer just have to decide \emph{if} there is a need for a control adjustment, but also \emph{what type} of adjustment (and, in the just speculated case of a constant-rate primitive, whether to stop generating it, or switch to another rate). Such decisions could be modelled as competitions between accumulators \citep[cf.~e.g.,][]{UsherAndMcClelland2001, PurcellEtAl2012} representing the different adjustment types, or in the active inference framework \citep{FristonEtAl2012DopamineAffordance} as competing predictions of what type of control will be carried out next. Indeed, even the present single-primitive formulation of our framework could be extended in this direction, by casting the individual amplitudes of stepwise position change as competing decisions \citep[cf.~e.g.,][]{ErlhagenAndSchoner2002, Cisek2007}\footnote{One specific benefit of such an approach would be that it would allow sensory noise to affect both timing and amplitude of adjustments, whereas the present framework decouples sensory and motor noise completely.}.

\subsection{Open-loop versus closed-loop, intermittent versus continuous}
\label{sec:OpenLoopVsClosedLoopDiscussion}
As has been mentioned above, it remains contentious whether, and if so to what extent and in what types of tasks, the nervous system engages in intermittent control, and \cite{GawthropEtAl2011} have argued that part of the empirical difficulty might lie in the capability of intermittent controls to \inquotes{masquerade} as continuous control. 
Another, related theme in the literature has been that the nervous system might be capable of combinations of open-loop and closed-loop control, and/or of continuous and intermittent control. Such hybrid control can be achieved for example by intermittently turning a continuous controller on and off \citep{CollinsAndDeLuca1993, AsaiEtAl2009}, by following up an open-loop primitive with a period of continuous closed-loop control \citep{MartinezGarciaEtAl2016}, or by applying \emph{system-matched holds} which are open-loop but continuous and highly flexible to be optimal with respect to the controlled system and situation \citep{GawthropEtAl2011}.

In making these types of distinctions, to not exaggerate the theoretical disagreement it seems important to be careful about what is meant by the terms being used, and at what level of analysis. As has already been discussed above, an action which is open-loop and ballistic at one level of a control hierarchy (e.g., a control adjustment of amplitude $\tilde{g}$ triggered in response to a prediction error $\epsilon$, but unaffected by later changes in $\epsilon$) might be implemented in closed-loop control at a lower level (e.g., ensuring that the performed amplitude is actually $\tilde{g}$; which again might rely on open-loop bursts of movement at an even lower, spinal level). Furthermore, higher up in the hierarchy the open-loop action might be part of a more sustained behaviour which is closed-loop in nature (e.g., a sequence of open-loop adjustments with amplitudes $\tilde{g}_i$, each in well-tuned response to the $\epsilon$ at time of adjustment onset). Something similar holds for the distinction between continuous and intermittent control; movement within an individual kinematic primitive is certainly continuous, and sequences of superpositioned intermittent kinematic primitives can generate continuous movement of arbitrary nature. 

Even with the above clarification, there can of course still be disagreement about whether, at a given level of analysis, sensorimotor control is best described as closed-loop or open-loop, continuous or intermittent. These discussions are probably best held at a task-specific level, with support from task-specific evidence. Hopefully the task-general framework proposed in this article can provide some useful inputs to such work.

However, one task-general counter-question that could be asked in response to the hybrid control schemes mentioned above, is whether the hypothesised episodes of continuous and (by some accounts) closed-loop behaviour could not again be instances of intermittent control masquerading as continuous? As suggested in the section just above, such a masquerade could come not only in the form of a succession of motor primitives triggered in closed loop, but also possibly as a learned, open-loop sequence of simpler primitives, superpositioned to construct a more complex motor action (e.g., to implement a system-matched hold). To clarify these matters, one would first need to locate candidates for the hypothesised episodes of hybrid control, and then subject them to detailed investigation. 




\subsection{Sensory prediction from corollary discharge primitives}

As already touched upon, the idea of prediction (or more specifically \emph{predictive coding} or \emph{predictive processing}) is much emphasised in many contemporary accounts of perception, cognition, and action \citep{RaoAndBallard1999, Friston2005, Friston2010, Clark2013, Clark2016, Hohwy2013, EngstromEtAl2017}. As discussed above at several places in this article, many previous authors have also highlighted the specific importance for sensorimotor control of Smith Predictor-like mechanisms, but to our knowledge this has so far always been in the context of continuous control. Here, we have integrated such a mechanism into an intermittent control framework, where it arguably is even more useful\footnote{Another type of prediction, which has been used in previous intermittent control models, addresses specifically motor delays by basing the control not on the current control error (or control error prediction error), but instead on what the error will have become by the time the motor action gets effectuated \citep[e.g., ][]{GawthropEtAl2009, GawthropEtAl2011, GordonAndSrinivasan2014}. In our understanding this type of prediction differs from the corollary discharge, \inquotes{sensory consequence of motor actions} type of prediction for which there is ample neuroscientific support, as reviewed in \secref{PredictionConcept}. However, it should be possible to incorporate also this type of prediction in the present framework, as part of the definition of the quantity $P$.}. 

The other, and possibly more important, theoretical contribution of this paper with respect to prediction, is the insight that a useful prediction signal can be constructed by superposition of simple \inquotes{prediction primitives}, triggered in parallel with each new control adjustment. As mentioned above, when these prediction primitives are mathematically derived to be (near) optimal for the tasks studied here (manual tracking and car steering), they obtain a shape that is similar in nature to corollary discharge biases that have been recorded in for example crickets and electric fish \citep{PouletAndHedwig2007, ChagnaudAndBass2013, RequarthAndSawtell2014}. These recorded corollary discharges have also been shown to change in shape with the motor action that triggers them \citep{ChagnaudAndBass2013, RequarthAndSawtell2014}, just as $H$ has been suggested to depend on $G$ here, and repeated corollary discharges are summed on top of each other in a fashion that is reminiscent of linear superposition \citep{ChagnaudAndBass2013}. In other words, the present computational formulation of sensory prediction could possibly map very directly onto actual neural mechanisms and signals.

If so, this suggests a heuristic strategy for the construction of forward model transfer functions, where an isolated corollary discharge or prediction primitive is somewhat analogous to the step response of the system (or the response to whichever motor primitive in question), at the level of the expressed controlled perceptual quantity. It should be pointed out, however, that this might typically be a rather approximate forward model, due to the prediction primitive itself only being an approximate step response, such as proposed here for the car steering task, and/or due to the controlled system not reacting to superpositioned motor inputs in exactly the same ways as the corollary discharges get summed together. Furthermore, there is also a difference from a typical forward model in control theory, in how the prediction primitive here first \inquotes{resets} the prediction error signal to the currently observed prediction error, then falls from there.

Given the above theoretical arguments, it is encouraging that our behavioural observations here aligned with the idea that prediction error determines control adjustment amplitudes; car steering adjustment amplitudes were better explained as a linear scaling of the prediction error $\epsilon = \Prec - \Pp$ than as a linear scaling of $\Prec$ directly. 
A possible alternative interpretation of this finding, without involving sensory prediction, would be that the secondary, corrective adjustments which were better explained by $\epsilon$ than by $\Prec$, were actually not triggered in response to either of these, but instead as part of a longer-duration, multi-step open-loop primitive, e.g.~to change yaw angle rather than yaw rate (as discussed in \secref{PrimitiveTypesDiscussion}). Visual inspection tentatively speaks against that hypothesis, at least in its simplest form; for example, the pair of stepwise adjustments starting at 0.4 s and 0.8 s in \figsref{ExampleIntermittentReconstructions}(c) and \ref{fig:ExampleAmplitudeFits}(c) affects both yaw angle and yaw rate. In any case, these findings and alternative hypotheses deserve to be followed up in more detail in targeted experiments, e.g.~with better control of the errors being responded to at control adjustment onset.

Such experiments could also look closer at the mentioned difference between drivers, with the prediction-based amplitude model providing a slightly worse fit than the prediction-free version for some drivers. It should be investigated whether these are random fluctuations in the data, or perhaps an indication of differences in control strategy between individuals.


\subsection{Near-optimal control of percepts versus optimal control of a system}
As was mentioned in the Introduction, several accounts have described sensorimotor control as an optimal control of the body and its environment \citep[e.g., ][]{KleinmanEtAl1970, TodorovAndJordan2002, ShadmehrAndKrakauer2008, FranklinAndWolpert2011, GawthropEtAl2011}, whereas others have suggested that it might be misleading to make too strong analogies between the nervous system and optimal controllers such as designed by engineers \citep{Friston2011, PickeringAndClark2014}. The framework proposed here aligns with the latter view, and also provides a concrete suggestion for how the nervous system might achieve near-optimal sensorimotor control by a careful combination of mechanisms which are all in themselves ad hoc and approximate in nature: perceptual heuristics, noisy evidence accumulation, a limited set of predefined motor primitives, and approximate but sufficiently effective sensory predictions. 

It should be emphasised that there is a sense in which these two accounts are very compatible; if regarded as another case of models operating at different levels of description. A non-strict interpretation of the optimal control type of account--which for example \cite{TodorovAndJordan2002} seem to support--is that it is at its most useful at a purely behavioural level, for well-practised tasks where the nervous system has been able to learn how to achieve something close to optimal control. At this level of description, engineering-type optimal control has proven powerful as a tool for predicting what behaviour might look like under a wide variety of tasks. 

In contrast, the type of framework proposed here will typically need more meticulous, task-specific attention, for example to identify and parameterise the relevant perceptual heuristics, before good predictions about behaviour can be made. On the other hand, if the present framework does indeed provide a more accurate description of the actual mechanisms involved, it should lend itself better to various forms of generalisation. For example, once properly established, the type of model proposed here might provide more accurate predictions of how sensorimotor behaviour generalises to novel situations \citep[important not least in a driving context; ][]{Markkula2014, Markkula2015}. Furthermore, models that are based on appropriate component mechanisms should in principle also be more suitable as starting points for accounts of how various factors influence sensorimotor control. Here, extra leverage can be had from the large neuroscientific literatures about the various component mechanisms; there is for example existing knowledge about how evidence accumulation processes might accomodate multisensory integration \citep{NoppeneyEtAl2010, RaposoEtAl2012}, and how they are affected by variations in arousal \citep{JepmaEtAl2009, RatcliffAndVanDongen2011} or time pressure \citep{BogaczEtAl2010}.

\subsection{Applying the framework to other sensorimotor tasks}

If one should wish to apply the present framework to other sensorimotor tasks beyond what has been studied here, the most obvious candidates would be tasks that are similar in nature to car steering; visuo-manual control tasks involving some external plant with dynamics of its own. Such tasks include laboratory-type joystick tracking tasks, robotic teleoperation in medicine or space, and longitudinal, lateral, and vertical control of other types of vehicles than cars; on land, in air, or on sea. In these tasks, novel and useful modelling could possibly be done with minimal or no modification to the computational framework presented here.

As already hinted above, such immediate applicability seems less likely for sensorimotor tasks which have been deeply investigated in the laboratory, such as ocular tracking, manual reaching, and postural control. In these contexts, it might nevertheless be useful to consider the adoption, into existing task-specific models, of some of the component mechanisms suggested here. For example, might evidence accumulation mechanisms help explain better the timing of catch-up saccades during smooth pursuit eye movements \citep{DeBrouwerEtAl2002, GrossbergEtAl2012}, of leg muscle activation in quiet standing, or of corrective submovements in reaching?  Could it be beneficial to model the apparent intermittency of postural balance control as stereotyped motor primitives rather than episodes of continuous control \citep{AsaiEtAl2009} or system-based holds \citep{GawthropEtAl2011}, and what about introducing a Smith Predictor control scheme in models of these tasks? And is it completely clear that \cite{Westheimer1954} was wrong in suggesting that smooth pursuit eye movements are constructed from something like the control rate primitives speculated in \secref{PrimitiveTypesDiscussion}?

In some of the tasks mentioned above, it might be desirable to consider the types of mechanisms discussed here in an expanded hierarchy with several levels, as briefly discussed in \secref{OpenLoopVsClosedLoopDiscussion}.

\section{Conclusion}
\label{sec:Conclusion}

It has been proposed, here, that intermittent sensorimotor control is achieved by the nervous system as ballistic motor primitives triggered after accumulation to threshold of errors in prediction of perceptual quantities indicating the need for control (control error prediction errors). These ideas have been realised in a computational framework for the special case of one-dimensional stepwise control, 
and it has been shown how existing models based on one-dimensional continuous control laws can be generalised to intermittent control using this framework.
Such generalisation has been demonstrated by formulation of one simple example model of a manual tracking task, and a more complete example of car steering control.

With the assumptions of the framework as a starting point, and supported by a simple method for interpreting a control signal as intermittent control, two data sets of human car steering have been analysed.
The results show that the observed steering lends itself well to being understood as a sequence of sigmoidal step adjustments, the amplitudes of which can be explained using an existing, originally continuous, model.
The fit of this amplitude model is further improved if assuming that what the drivers respond to is not the error-describing perceptual quantity itself, but rather errors in prediction of this quantity.
Furthermore, distributions of control adjustment timing, and how these covary with adjustment amplitude, were seemingly better explained by a model assuming evidence accumulation instead of mechanisms typical of existing intermittent control models (error deadzones and minimum refractory periods).
More targeted empirical work, in both driving and other sensorimotor tasks, is warranted to verify the findings presented here, especially those relating to the possible roles of evidence accumulation and sensory prediction in intermittent sensorimotor control.

A novel theoretical insight, here, has been that not only the motor output but also the sensory prediction can be usefully constructed from a superposition of discrete primitives, to yield a prediction signal that might not be exact but accurate enough for successful behaviour. Interestingly, the nature of this type of prediction signal, as suggested for the tasks studied here, is reminiscent of corollary discharge biases as observed in animals. This could provide another piece of the puzzle in the debate regarding to what extent and how the nervous system might act as an optimal controller.

The present account aligns with the general idea, and suggests a concrete computational realisation of it, that a number of mechanisms that are all approximate and ad hoc in nature (ballistic motor primitives, perceptual heuristics, noisy evidence accumulation, corollary discharge prediction primitives) are used in concert by the nervous system to achieve behaviour that is near-optimal under a wide range of circumstances.

In sum, the presently proposed framework provides an intermediate-level, behavioural account of sensorimotor control, by integrating, conceptually and computationally, a set of neurobiologically plausible mechanisms that have been present in isolation in previous models. The closer connection to neurobiology could be preferable to the optimal control level of description in some contexts, and the task-general ideas outlined here could provide interesting directions for future development of more detailed task-specific models.

\section*{Data statement}
Upon publication of this paper, the research data supporting it will be made publicly available.


\bibliographystyle{spbasic}      
\bibliography{Bibliography_BiolCyb_Markkula}   

\end{document}